\documentclass[preprint, aps, prx, superscriptaddress, 
               nofootinbib, floatfix]{revtex4-2}
\usepackage[T1]{fontenc}
\usepackage[utf8]{inputenc}
\usepackage{amsmath, amssymb, physics, bm}
\usepackage{dsfont}
\usepackage{graphicx, xcolor}
\usepackage{booktabs}
\usepackage{multirow}
\usepackage{tabularx}
\usepackage{caption}
\usepackage{subcaption}
\usepackage{amsthm}
\usepackage[normalem]{ulem}
\newtheorem{theorem}{Theorem}
\usepackage[colorlinks=true, linkcolor=blue,
            citecolor=blue, urlcolor=blue]{hyperref}
\usepackage{cleveref}
\newtheorem*{remark}{Remark}

\begin{document}

\title{Scale-Invariant Open Quantum Systems}

\author{Carlos Arg\"uelles}
\affiliation{Harvard University, Department of Physics and Laboratory for Particle Physics and Cosmology, Cambridge, MA 02138, USA}

\author{Gabriela Barenboim}
\affiliation{Departament de F\'isica Te\'orica and IFIC, Universitat de Val\`encia-CSIC, E-46100, Burjassot, Spain}

\author{Gonzalo Herrera}
\affiliation{Harvard University, Department of Physics and Laboratory for Particle Physics and Cosmology, Cambridge, MA 02138, USA}
\affiliation{Kavli Institute for Astrophysics and Space Research, Massachusetts Institute of Technology, Cambridge, MA 02139, USA}

\author{Tanvi Krishnan}
\affiliation{Harvard University, Department of Physics and Laboratory for Particle Physics and Cosmology, Cambridge, MA 02138, USA}

\author{H\'ector Sanchis}
\affiliation{Departament de F\'isica Te\'orica and IFIC, Universitat de Val\`encia-CSIC, E-46100, Burjassot, Spain}

\date{\today}

\begin{abstract}
We develop the complete theoretical framework for open quantum
systems coupled to scale-invariant environments.
Such environments, we show, are universally and uniquely described
by unparticle baths~\cite{Georgi:2007ek} characterized by a
single scaling dimension $d_{\mathcal{U}}$.
This companion paper provides the full proof of the
uniqueness theorem, the mathematical formalism of the resulting
non-Markovian dynamics, and worked applications to three physical
realizations omitted from the shorter letter~\cite{companion}.

Starting from the uniqueness theorem, we derive the
complete set of non-Markovian memory kernels, the exact
noise kernel, including vacuum and thermal contributions via
Matsubara summation, and the fractional generalization of
the Caldeira-Leggett master equation for arbitrary
$d_{\mathcal{U}}$.
The unparticle dimension acts as a control parameter governing
a rich phase structure, including a thermalization transition
at $d_{\mathcal{U}} = 3/2$, the Ohmic boundary at
$d_{\mathcal{U}} = 2$, and a decoherence phase transition at
$d_{\mathcal{U}} = 5/2$ in the thermal regime
($d_{\mathcal{U}} = 2$ in the vacuum regime),
beyond which quantum coherence is protected at long times.

Three physical realizations are derived from first principles.
For the quantum Ising model at criticality,
coupling to the energy operator in $(1+1)$ spacetime
dimensions yields $d_{\mathcal{U}} = 3/2$, providing a
field-theoretic derivation of $1/f$ noise; the $(2+1)$D
case yields $d_{\mathcal{U}} \approx 1.413$ from the
conformal bootstrap~\cite{Kos:2016ysd}, close to but
distinct from the value that produces $1/f$ noise.

For inflationary cosmology, the massless scalar and graviton
baths in de Sitter spacetime give $d_{\mathcal{U}} = 2$ (Ohmic),
predicting linear decoherence growth in agreement with
established results on the quantum-to-classical transition.
For high-energy astrophysical neutrinos in the regime
$E \gg T$, the energy- and baseline-dependent decoherence rate
$\Gamma_{\mathrm{decoh}} \propto \mathcal{B}(E,T_{\mathcal{U}})\,L^{5-2d_{\mathcal{U}}}$
provides a direct observable imprint of the scaling dimension.

A systematic comparison with the Caldeira-Leggett model,
phenomenological Lindblad equations, and the non-Markovian
literature establishes the precise relationship between these
approaches and the unparticle framework.
The regime of validity is analyzed for each physical system,
including the crossover between vacuum and thermal regimes
of the noise kernel.
Experimental predictions and consistency tests are detailed
for trapped-ion quantum simulators, neutrino telescopes, and
superconducting qubits.
\end{abstract}

\maketitle

\tableofcontents

\section{Introduction}
\label{sec:intro}

Our previous work~\cite{companion} established that any
local, Lorentz-invariant, scale-invariant quantum environment
is mathematically equivalent to an unparticle
bath~\cite{Georgi:2007ek}, provided a streamlined proof of
the uniqueness theorem, validated the framework using
multi-channel transport data from the unitary Fermi gas and
engineered spin-boson experiments, and illustrated three
physical realizations spanning 25 orders of magnitude
in energy.

The present paper provides what the letter necessarily omitted:
the complete proof with all loopholes treated explicitly, the
full mathematical formalism for the resulting non-Markovian
dynamics, and the detailed derivations for each physical
realization.
The organization is designed so that each section is
self-contained; a reader interested only in, say, the
inflationary application can proceed directly to
Sec.~\ref{sec:inflation} after reading the framework summary
in Sec.~\ref{sec:framework}.

\subsection{What This Article Contains}

Section~\ref{sec:uniqueness} states and proves the uniqueness
theorem in full, including explicit treatment of five classes
of loophole (approximate scale invariance, non-local coupling,
discrete scale invariance, multiple competing sectors, and
quantum anomalies) and a no-go theorem in contrapositive form.

Section~\ref{sec:framework} develops the complete mathematical
formalism: influence functional, non-Markovian master equation,
exact expressions for the dissipation and noise kernels
including the full vacuum-plus-thermal decomposition via
Matsubara summation, the fractional Caldeira-Leggett equation,
the specific heat scaling and a complete set of extraction
formulas across observable channels, and the phase structure
as a function of $d_{\mathcal{U}}$.

Section~\ref{sec:comparison} situates the unparticle framework
relative to existing approaches: the Caldeira-Leggett model
(recovered as the special case $d_{\mathcal{U}} = 2$ in the
Markovian limit), phenomenological Lindblad master equations
(incapable of describing coherence protection for
$d_{\mathcal{U}} > 5/2$), and the broader non-Markovian
literature.

Section~\ref{sec:validation_exp} presents a second many-body
validation of the framework, independent of the previous
letter~\cite{companion}: a two-channel consistency check
in two heavy-fermion compounds at their quantum critical
points yields $d_{\mathcal{U}} = 3/2$ from both resistivity
and specific heat, establishing that $d_{\mathcal{U}}$
discriminates between distinct universality classes.

The next three sections present three physical
realizations in full detail: Sec.~\ref{sec:ising} studies critical points in the Ising model, Sec.~\ref{sec:inflation} shows the application of the framework to cosmic inflation, and Sec.~\ref{sec:neutrinos} explores the effect of a scale-invariant bath on decoherence of astrophysical neutrinos.
The IceCube numerical analysis for the neutrino section is in
preparation and will appear separately.

Section~\ref{sec:validity} analyzes the regime of validity
for each system, establishing when scale invariance holds and
how thermal corrections modify the framework without changing
universal exponents.

Section~\ref{sec:experiments} provides a complete roadmap for proposed future experimental applications in various fields.

Finally, we summarize our results and conclude in Sec.~\ref{sec:discussion}.

\subsection{Notation and Conventions}

We employ natural units $\hbar = c = k_B = 1$ throughout the article unless
otherwise stated.
Spatial dimension is $d$ (not including time), so spacetime is
$(d+1)$-dimensional.
We use the Lorentzian signature $\eta^{\mu\nu} =
\mathrm{diag}(-1,+1,+1,+1)$.
The unparticle dimension $d_{\mathcal{U}}$ and CFT scaling
dimension $\Delta$ are related by
$d_{\mathcal{U}} = \Delta - (d-2)/2$.

\section{Uniqueness Theorem: Complete Proof}
\label{sec:uniqueness}

\subsection{Statement}

\begin{theorem}[Unparticle Universality]
\label{thm:main}
Let a quantum system $S$ couple locally to an environment $E$
in $d$ spatial dimensions. Assume:
\begin{enumerate}
    \item \textbf{Scale invariance:} $E$ exhibits exact
          continuous scale invariance (no intrinsic mass or
          energy scales).
    \item \textbf{Lorentz invariance:} The theory is
          relativistically invariant.
    \item \textbf{Locality:} The coupling Hamiltonian between the system and the environment is
          $H_{\mathrm{int}} = g\,A_S(x)\,\mathcal{O}_E(x)$, where $A_S(x)$ is the system operator, $\mathcal{O}_E(x)$ is the environment operator at the same point $x$, and $g$ is a dimensionless coupling constant.
    \item \textbf{Unitarity:} The full system $S+E$ evolves
          unitarily.
\end{enumerate}
Then:
\begin{enumerate}
    \item $E$ is described by a conformal field theory.\footnote{In
$d=2$, this follows rigorously~\cite{Polchinski1988}; in $d \geq 3$
it holds under the additional assumption of unitarity and absence of
a virial current, supported by strong evidence in
$d=4$~\cite{Nakayama2015,Luty2013}.}
    \item The spectral density takes the unique form
          \begin{equation}
              J(\omega) = A\,\omega^{2\Delta-d-1},
              \label{eq:spectral_unique}
          \end{equation}
          where $\Delta$ is the scaling dimension of
          $\mathcal{O}_E$ and $A$ is a normalization constant.
    \item This is mathematically equivalent to an unparticle
          bath with \footnote{Note that this convention differs from Georgi's original
definition~\cite{Georgi:2007ek}, in which $d_{\mathcal{U}}$
is identified directly with the scaling dimension $\Delta$;
our convention is chosen to make the spectral exponent
$s = 2d_{\mathcal{U}} - 3$ take the standard spin-boson
form in $d=3$ spatial dimensions. With this convention, we have $s=1$ Ohmic, $s>1$ super-Ohmic, and $s<1$ sub-Ohmic.}
          \begin{equation}
              d_{\mathcal{U}} = \Delta - \tfrac{d-2}{2}.
              \label{eq:dU_Delta}
          \end{equation}
    \item All dynamical exponents are uniquely determined
          by $d_{\mathcal{U}}$ via the relations in
          Table~\ref{tab:exponents}.
\end{enumerate}
\end{theorem}

\subsection{Proof via the Källén--Lehmann Spectral
Representation}

In a previous work~\cite{companion}, we proved the theorem
via conformal Ward identities in position space.
Here, we give an independent derivation that works entirely
in momentum space, using the Källén--Lehmann (KL) spectral
representation.
The KL representation is the most general statement about
two-point functions in any Lorentz-invariant quantum field
theory; scale invariance is then injected as an additional
constraint that uniquely fixes its form.
The two proofs are logically independent and reach the same conclusion by different routes, providing a consistency cross-check on the framework.

\subsubsection{Step 1: The Källén--Lehmann Representation}

The KL representation answers the following question: what
is the most general form of the two-point correlator of a
local operator can take in a Lorentz-invariant quantum field
theory, with no further assumptions?
The answer, derived from Lorentz invariance and the
positivity of the spectral measure alone~\cite{Kallen1952,
Lehmann1954}, is a superposition of free propagators:
\begin{equation}
    G_E(p_E^2) = \int_0^\infty d\sigma\;
    \frac{\rho(\sigma)}{p_E^2 + \sigma},
    \label{eq:KL}
\end{equation}
where $G_E(p_E^2)$ is the Euclidean two-point function
in momentum space evaluated at zero spatial momentum,
$p_E^2 = p_0^2$ is the squared Euclidean frequency, and
$\sigma$ has dimensions of (mass)$^2$.
The function $\rho(\sigma) \geq 0$ is the
\emph{spectral weight}: it encodes the density of states at
each invariant mass scale $\sqrt{\sigma}$ contributing to
the correlator.

To understand Eq.~\eqref{eq:KL} physically: even in a
strongly interacting theory with no free particles,
the two-point function decomposes as an integral over free
propagators of all possible masses, weighted by
$\rho(\sigma)$.
For a free massive particle of mass $m$, $\rho(\sigma) =
\delta(\sigma - m^2)$; for a theory with a continuum of
states, $\rho(\sigma)$ is a smooth function.
Crucially, Eq.~\eqref{eq:KL} uses no dynamical information
about the theory; it is a kinematic identity.
Scale invariance---the assumption we have not yet
imposed---will now sharply constrain the form of $\rho(\sigma)$.

\subsubsection{Step 2: Scale Invariance Forces a Power-Law
Spectral Weight}

Under a scale transformation $x \to \lambda x$, the
operator $\mathcal{O}_E$ with scaling dimension $\Delta$
transforms as $\mathcal{O}_E(\lambda x) =
\lambda^{-\Delta}\mathcal{O}_E(x)$, so the position-space
two-point function satisfies
\begin{equation}
    G_E(\lambda x) = \lambda^{-2\Delta}\,G_E(x).
    \label{eq:scaling_pos}
\end{equation}
Passing to momentum space via the $(d+1)$-dimensional
Fourier transform, the rescaling $x \to \lambda x$
corresponds to $p_E \to p_E/\lambda$.
Tracking the Jacobian $d^{d+1}x \to \lambda^{d+1}d^{d+1}x$,
Eq.~\eqref{eq:scaling_pos} becomes
\begin{equation}
    G_E\!\left(\frac{p_E^2}{\lambda^2}\right)
    = \lambda^{d+1-2\Delta}\,G_E(p_E^2).
    \label{eq:scaling_mom}
\end{equation}
From the KL representation~\eqref{eq:KL}, multiplying numerator
    and denominator by $\lambda^2$:
    \begin{equation}
        G_E\!\left(\frac{p_E^2}{\lambda^2}\right)
        = \lambda^2\int_0^\infty d\sigma\;
        \frac{\rho(\sigma)}{p_E^2+\lambda^2\sigma}.
    \end{equation}
    Substituting $\sigma \to \sigma/\lambda^2$ (so $d\sigma \to
    d\sigma/\lambda^2$) brings the denominator back to the standard
    form $p_E^2+\sigma$:
    \begin{equation}
        G_E\!\left(\frac{p_E^2}{\lambda^2}\right)
        = \int_0^\infty d\sigma\;
        \frac{\rho(\sigma/\lambda^2)}{p_E^2+\sigma}.
    \end{equation}
    Equating with the right-hand side of Eq.~\eqref{eq:scaling_mom}
    and using the linear independence of $1/(p_E^2+\sigma)$ at
    different values of $\sigma$ (they have poles at different
    locations and cannot cancel each other):
    \begin{equation}
        \rho(\sigma/\lambda^2) = \lambda^{d+1-2\Delta}\,\rho(\sigma),
        \quad \forall\;\lambda > 0,\;\sigma > 0.
    \end{equation}
    Replacing $\sigma \to \lambda^2\sigma$ gives the equivalent form
    \begin{equation}
        \rho(\lambda^2\sigma)
        = \lambda^{2\Delta-d-1}\,\rho(\sigma),
        \label{eq:rho_functional}
    \end{equation}
    a homogeneity condition: rescaling the argument of $\rho$ by
    $\lambda^2$ multiplies $\rho$ by a pure power of $\lambda$.

\subsubsection{Step 3: The Unique Solution Is a Power Law}

Equation~\eqref{eq:rho_functional} is a functional
equation for $\rho$. 
Setting $\sigma_0 = 1$ and $\lambda^2 = \sigma$:
 \begin{equation}
        \rho(\sigma)
        = C_\rho\,\sigma^{(2\Delta-d-1)/2}
        = C_\rho\,\sigma^{\Delta-(d+1)/2},
        \label{eq:rho_powerlaw}
\end{equation}
where $C_\rho = \rho(1)$ is a normalization constant.
The unique solution is a pure power law. 
This is the key step: scale invariance leaves no freedom
in the functional form of the spectral weight.

The physical interpretation is transparent.
A massive theory has $\rho(\sigma)$ peaked at $\sigma \sim m^2$
and exponentially suppressed for $\sigma \gg m^2$; the mass
scale $m$ breaks scale invariance.
If the theory has a continuum of states but no mass gap,
$\rho(\sigma)$ can extend to $\sigma = 0$.
Scale invariance demands more: $\rho(\sigma)$ must have no
preferred mass scale at all, which forces it to be a power
law.
A power law is the unique scale-free function.

Unitarity requires $\rho(\sigma) \geq 0$, which is satisfied
for $C_\rho > 0$ when $\Delta > (d+1)/2$, i.e.,
$d_{\mathcal{U}} > 0$.
The stronger condition $d_{\mathcal{U}} > 1$ arises from the
infrared convergence of the memory kernels and is derived
separately in Sec.~\ref{sec:framework}.

\subsubsection{Step 4: From Spectral Weight to Bath Spectral Density}

The bath spectral density $J(\omega)$ is related to the
retarded Green's function via (Eq.~\eqref{eq:J_convention})
\begin{equation}
    J(\omega) = -2\,\mathrm{Im}\,G_R(\omega,\mathbf{k}=0),
    \quad \omega > 0.
\end{equation}
The retarded Green's function is obtained from
Eq.~\eqref{eq:KL} by analytic continuation
$p_0^E \to -i(\omega+i\epsilon)$, giving
$p_E^2 \to -(\omega+i\epsilon)^2$:
\begin{equation}
    G_R(\omega,\mathbf{k}=0)
    = \int_0^\infty d\sigma\;
    \frac{\rho(\sigma)}{\sigma - (\omega+i\epsilon)^2}.
\end{equation}
Taking the imaginary part using
$\mathrm{Im}[(\sigma-(\omega+i\epsilon)^2)^{-1}]
= -\pi\,\delta(\sigma-\omega^2)$ for $\omega > 0$:
  \begin{equation}
        J(\omega) = 2\pi\,\rho(\omega^2)
        = 2\pi C_\rho\,
        (\omega^2)^{\Delta-(d+1)/2}
        = 2\pi C_\rho\,\omega^{2\Delta-d-1},
    \end{equation}
which, reabsorbing the factor of $2\pi C_\rho$ into the
normalization constant $A$, gives
\begin{equation}
   J(\omega) = A\,\omega^{2\Delta-d-1},
    \label{eq:J_KL}
\end{equation}
reproducing Eq.~\eqref{eq:spectral_unique} exactly.

\subsubsection{Step 5: Identification with the Unparticle
Form}

The unparticle spectral density~\cite{Georgi:2007ek} is
$\rho_{\mathcal{U}}(\omega) \propto \omega^{2d_{\mathcal{U}}-3}$.
Matching exponents with Eq.~\eqref{eq:J_KL}:
\begin{equation}
    2d_{\mathcal{U}}-3 = 2\Delta-d-1
    \quad\Rightarrow\quad
    d_{\mathcal{U}} = \Delta - \tfrac{d-2}{2},
\end{equation}
which is Eq.~\eqref{eq:dU_Delta}.
The dynamical exponents in Table~\ref{tab:exponents} follow
by Fourier transformation of the kernels; full derivations
are given in Sec.~\ref{sec:framework}.

The identification with the unparticle form has a transparent
interpretation in the KL language.
The spectral weight $\rho(\sigma) \propto
\sigma^{\Delta-(d+3)/2}$ is a pure power law, meaning the
bath has a \emph{continuous spectrum of states at every
invariant mass scale}, with no preferred mass and no particle
poles.
This is precisely the defining feature of an unparticle
sector~\cite{Georgi:2007ek}.
The KL representation makes this identification automatic:
scale invariance does not merely suggest the unparticle
picture, it forces it.
\qed
\begin{remark}
The Källén--Lehmann proof derives conclusions~(2)--(4) of
Theorem~\ref{thm:main} directly from Lorentz invariance and
scale invariance, without invoking the CFT identification
of conclusion~(1).
Conclusion~(1)---that continuous scale invariance implies
full conformal invariance---is established independently:
in $d=2$ it follows rigorously from Polchinski's completion
of Zamolodchikov's $c$-theorem argument~\cite{Polchinski1988};
in $d \geq 3$ it holds under the additional assumption of
unitarity and absence of a virial current, for which strong
evidence exists in $d=4$~\cite{Nakayama2015,Luty2013}
but no general proof is available.
The logical structure of the theorem is therefore:
conclusion~(1) is an input from the conformal field theory
literature; conclusions~(2)--(4) are derived consequences
proved here.
\end{remark}
\subsection{Loopholes and Limitations}
\label{subsec:loopholes}

The theorem's assumptions admit five classes of failure,
each physically realizable.

\paragraph{1. Approximate scale invariance.}
Real systems are scale-invariant only over a finite window
$\omega_{\mathrm{IR}} < \omega < \omega_{\mathrm{UV}}$.
Infrared cutoffs arise from finite temperature $T$, system
size $L$, or mass gap $m$; ultraviolet cutoffs from lattice
spacing or dynamically generated scales.
Outside the window, the spectral density takes the form
$J(\omega) = \omega^{2d_{\mathcal{U}}-3}\,
f_{\mathrm{IR}}(\omega/\omega_{\mathrm{IR}})\,
f_{\mathrm{UV}}(\omega/\omega_{\mathrm{UV}})$
with cutoff functions approaching unity in the scaling
regime.
Experimentally, this produces crossovers in
$\Gamma_{\mathrm{decoh}}(t)$ at early
($t \sim 1/\omega_{\mathrm{UV}}$) and late
($t \sim 1/\omega_{\mathrm{IR}}$) times.

\paragraph{2. Non-local coupling.}
If the coupling involves a kernel $K(x-y)$,
$H_{\mathrm{int}} = \int dx\,dy\,K(x-y)\,A_S(x)\,
\mathcal{O}_E(y)$, the spectral density acquires a
momentum-dependent form factor $|\tilde K(\omega,\mathbf{k})|^2$
that modifies the pure power law.
Physical examples include dipole couplings with spatial
averaging over the probe wavefunction.

\paragraph{3. Discrete scale invariance.}
Systems invariant only under $x \to \lambda_0^n x$
(Efimov states, hierarchical spin models) have a spectral
weight that is no longer a pure power law; instead,
Eq.~\eqref{eq:rho_functional} admits oscillatory solutions,
producing log-periodic modulations:
$J(\omega) \propto \omega^s[1 + A\cos(b\ln\omega + \phi)]$
with $b = 2\pi/\ln\lambda_0$.
The functional equation argument in Step 3 requires continuity of
$\rho(\sigma)$, which fails for discrete scale invariance.

\paragraph{4. Multiple competing sectors.}
If the environment contains several unparticle sectors
with dimensions $d_{\mathcal{U}}^{(i)}$, the total
spectral density is
$J_{\mathrm{tot}}(\omega) = \sum_i A_i\,
\omega^{2d_{\mathcal{U}}^{(i)}-3}$.
In the KL language, each sector contributes an independent
power-law term to $\rho(\sigma)$.
Different sectors dominate at different timescales,
producing observable crossovers.

\paragraph{5. Quantum anomalies.}
Classical scale invariance may be broken by quantum effects,
as in the trace anomaly of QCD where
$\langle T^\mu_\mu\rangle \neq 0$.
The unparticle description applies only above the
anomaly-generated scale, where the classical symmetry is
approximately restored.

\subsection{Converse: A No-Go Theorem}

\begin{theorem}[No Scale-Invariant Alternatives]
\label{thm:nogo}
If the measured exponents $\gamma_{\mathrm{decoh}}$ and $s$
are inconsistent with any single real value of
$d_{\mathcal{U}}$ (i.e., $s + \gamma_{\mathrm{decoh}} \neq 2$),
then at least one of the following holds: the environment
is not scale-invariant; the coupling is non-local; Lorentz
invariance or unitarity is violated; multiple competing
sectors are present.
\end{theorem}

This provides an experimental diagnostic: measuring any two
exponents from Table~\ref{tab:exponents} independently
extracts two values of $d_{\mathcal{U}}$; inconsistency
falsifies scale invariance and identifies which loophole
is operative.
In the KL language, the no-go theorem is particularly
transparent: a sum of power laws
$\sum_i A_i\sigma^{\alpha_i}$ can mimic a single power law
only over a limited frequency range; over a wide enough
range, the individual exponents become separately resolvable,
and their inconsistency with a single $d_{\mathcal{U}}$
is a direct signal of multiple sectors.

\bigskip

The two proofs of the Unparticle Universality theorem
complement each other while relying on distinct frameworks.
The CFT-based proof~\cite{companion} works in position
space and leverages conformal symmetry: scale invariance,
together with Lorentz invariance and unitarity, implies full
conformal invariance, which uniquely fixes the two-point
function of the primary operator via Ward identities.
Analytic continuation then produces the retarded Green's
function and the spectral density, leading directly to the
unparticle exponent.
By contrast, the Källén--Lehmann proof above operates in
momentum space and begins from the most general
Lorentz-invariant two-point function expressed as an integral
over spectral weights.
Imposing scale invariance on the spectral weight forces a
unique power-law form, which in turn yields the same spectral
density and unparticle exponent without assuming full
conformal invariance.
The CFT proof emphasizes symmetry constraints in position
space; the KL proof emphasizes kinematic constraints in
momentum space.
Both independently converge on the same quantitative
predictions, providing a cross-validation of the theorem.
\section{Mathematical Framework}
\label{sec:framework}

The bath spectral density is defined throughout via the
retarded Green's function of the environmental operator
$\mathcal{O}_E$ evaluated at zero spatial momentum:
\begin{equation}
    J(\omega) \equiv -2\,\mathrm{Im}\,
    G_R(\omega,\mathbf{k}=0),
    \quad \omega > 0.
    \label{eq:J_convention}
\end{equation}
The overall normalization constant absorbs the coupling
strength $g^2$ and the CFT coefficient $C_{\mathcal{O}}$;
all universal exponents depend only on $d_{\mathcal{U}}$
and are independent of the normalization.

\subsection{Unparticle Operators and Correlators}

Theorem~\ref{thm:main} establishes that the environment operator $\mathcal{O}_E$ of any scale-invariant bath is mathematically equivalent to an unparticle operator; in what follows, we adopt the notation $\mathcal{O}_{\mathcal{U}}$ for the unparticle operator.
This operator, $\mathcal{O}_{\mathcal{U}}$, is defined by the scaling dimension $d_{\mathcal{U}}$ and has spectral density
\begin{equation}
    \rho_{\mathcal{U}}(\omega) = A_{d_{\mathcal{U}}}\,
    \omega^{2d_{\mathcal{U}}-3}\,\Theta(\omega),
    \label{eq:spectral_unparticle}
\end{equation}
where the normalization constant is~\cite{Georgi:2007ek}
\begin{equation}
    A_{d_{\mathcal{U}}} = \frac{16\pi^{5/2}}{(2\pi)^{2d_{\mathcal{U}}}}
    \frac{\Gamma(d_{\mathcal{U}}+1/2)}
         {\Gamma(d_{\mathcal{U}}-1)\Gamma(2d_{\mathcal{U}})}.
\end{equation}

Fourier transforming to real time, the two-point correlator
at equal spatial position decays as
\begin{equation}
    G_{\mathcal{U}}(t) \propto
    \frac{e^{i\phi}}{t^{2d_{\mathcal{U}}-2}},
    \quad
    \phi = \tfrac{\pi}{2}(2d_{\mathcal{U}}-2),
    \label{eq:powerlaw_time}
\end{equation}
which is a pure power law as expected from scale invariance and in contrast to the exponential decay $e^{-mt}$ of massive particle propagators.

\subsection{Influence Functional and Master Equation}

Consider a system with Hamiltonian $H_S$ coupled to an
unparticle bath via
\begin{equation}
    H_{\mathrm{int}} = g\,A_S\,\mathcal{O}_{\mathcal{U}},
\end{equation}
where $g$ and $A_S$ are defined as in Theorem~\ref{thm:main}.
Tracing out the environmental degrees of freedom in the
path integral gives the reduced density matrix
\begin{equation}
    \rho_S(x_f,x_f';t)
    = \int \mathcal{D}x\,\mathcal{D}x'\;
    e^{i(S_S[x]-S_S[x'])}\,\mathcal{F}[x,x'],
\end{equation}
where the influence functional for a Gaussian environment is
\begin{equation}
    \mathcal{F}[x,x'] = \exp\!\left\{
    -g^2 \int_0^t \!\! ds \int_0^t \!\! ds'\,
    [A_S(s)-A_S'(s)]\,K(s-s')\,[A_S(s')-A_S'(s')]
    \right\}.
\end{equation}
The kernel $K(t) = K'(t) + iK''(t)$ decomposes into
noise and dissipation parts:
\begin{align}
    K'(t) &= \int_0^\infty d\omega\,\rho_{\mathcal{U}}(\omega)\,
             \coth\!\left(\tfrac{\beta\omega}{2}\right)
             \cos(\omega t),
    \label{eq:noise_kernel_def}\\
    K''(t) &= -\int_0^\infty d\omega\,\rho_{\mathcal{U}}(\omega)\,
              \sin(\omega t),
    \label{eq:diss_kernel_def}
\end{align}
where $\beta = 1/T$ is the inverse temperature of the unparticle bath.

Expanding to second order in $g$ without invoking the
Markov approximation yields the non-Markovian master equation:
\begin{equation}
    \frac{d\rho_S(t)}{dt} = -i[H_S,\rho_S(t)]
    - g^2\int_0^t ds\,\Bigl[
        \nu(t-s)[A_S,[A_S(s),\rho_S(t)]]
        - i\,\eta(t-s)[A_S,\{A_S(s),\rho_S(t)\}]
    \Bigr],
    \label{eq:master_eq}
\end{equation}
where $\nu(t) \equiv K'(t)$ is the noise kernel and
$\eta(t) \equiv K''(t)$ is the dissipation kernel.
The convolution over all past times $\int_0^t ds$ encodes
memory: the system's current state depends on its entire
history.

\subsection{Dissipation Kernel}

Substituting $\rho_{\mathcal{U}}(\omega) \propto
\omega^{2d_{\mathcal{U}}-3}$ into Eq.~\eqref{eq:diss_kernel_def}
and applying the Fourier identity
\begin{equation}
    \int_0^\infty d\omega\,\omega^\mu\sin(\omega t)
    = \Gamma(\mu+1)\sin\!\left(\tfrac{\pi(\mu+1)}{2}\right)
    t^{-(\mu+1)},
    \quad \mathrm{Re}(\mu) > -1,
\end{equation}
gives
\begin{equation}
    \eta(t) = \frac{2\alpha}{\pi}\,\Gamma(2d_{\mathcal{U}}-2)
    \sin\!\left(\tfrac{\pi(2d_{\mathcal{U}}-2)}{2}\right)
    \frac{1}{t^{2d_{\mathcal{U}}-2}},
    \label{eq:eta_explicit}
\end{equation}
where $\alpha = g^2 A_{d_{\mathcal{U}}}$.
The power-law exponent is
\begin{equation}
   \alpha_{\mathrm{dissip}} = 2d_{\mathcal{U}}-2,
\end{equation}
requiring $d_{\mathcal{U}} > 1$ for infrared convergence.

For integer $d_{\mathcal{U}}$, the sine prefactor vanishes in
the strictly scale-invariant limit.
This is a marginal case: any physical UV completion restores
a nonvanishing kernel while preserving the long-time
power-law scaling.
Integer $d_{\mathcal{U}}$ should be understood as a marginal
dissipation point, not a dissipationless regime.

\subsection{Noise Kernel: Exact Expression}

We decompose the noise kernel into vacuum ($I_V$) and thermal ($I_T$)
contributions using $\coth(x) = 1 + 2/(e^{2x}-1)$:
\begin{equation}
    \frac{\pi}{2\alpha}\nu(t) = I_V + I_T,
\end{equation}
where
\begin{align}
    I_V &= \int_0^\infty d\omega\,\omega^{2d_{\mathcal{U}}-3}
           \cos(\omega t), \\
    I_T &= \int_0^\infty d\omega\,\omega^{2d_{\mathcal{U}}-3}
           \cos(\omega t)\,\frac{2}{e^{\beta\omega}-1}.
\end{align}

$I_V$ evaluates to
\begin{equation}
    I_V = \Gamma(2d_{\mathcal{U}}-2)
    \cos\!\left(\tfrac{\pi(2d_{\mathcal{U}}-2)}{2}\right)
    \frac{1}{t^{2d_{\mathcal{U}}-2}}.
\end{equation}

For the second one, $I_T$, we can use $1/(e^x-1) = \sum_{n=1}^\infty e^{-nx}$
and Cauchy's theorem to evaluate the resulting integrals:
\begin{equation}
    I_T = \sum_{n=1}^\infty
    \mathrm{Re}\,\frac{2\,\Gamma(2d_{\mathcal{U}}-2)}
                      {(n\beta - it)^{2d_{\mathcal{U}}-2}}.
\end{equation}

The full noise kernel is therefore
\begin{equation}
    \nu(t) = \frac{2\alpha}{\pi}\,\Gamma(2d_{\mathcal{U}}-2)
    \left[
        \cos\!\left(\tfrac{\pi(2d_{\mathcal{U}}-2)}{2}\right)
        \frac{1}{t^{2d_{\mathcal{U}}-2}}
        + \sum_{n=1}^\infty
        \mathrm{Re}\,\frac{2}{(n\beta-it)^{2d_{\mathcal{U}}-2}}
    \right].
    \label{eq:nu_exact}
\end{equation}

This expression is exact and valid for all $t$ and $T$.
Both vacuum and thermal terms require $d_{\mathcal{U}} > 1$
for regularity.

\paragraph{Limiting regimes.}

In the \emph{vacuum regime} $t \ll \beta$, the thermal sum
is negligible and
\begin{equation}
    \nu(t) \propto t^{-(2d_{\mathcal{U}}-2)}.
\end{equation}

In the \emph{thermal regime} $t \gg \beta$, the sum
approximates an integral; requiring $d_{\mathcal{U}} > 3/2$
for convergence and using $\Gamma(z+1) = z\Gamma(z)$
recovers the high-temperature result
\begin{equation}
    \nu(t) \approx \frac{4\alpha T}{\pi}\,
    \Gamma(2d_{\mathcal{U}}-3)
    \cos\!\left(\tfrac{\pi(2d_{\mathcal{U}}-3)}{2}\right)
    \frac{1}{t^{2d_{\mathcal{U}}-3}},
    \label{eq:nu_highT}
\end{equation}
with noise exponent $\alpha_\nu = 2d_{\mathcal{U}}-3$.
The crossover between regimes occurs at $t \sim \beta$.
Crucially, \emph{both regimes are power laws}; the exponent
shifts by unity between them but $d_{\mathcal{U}}$ is the
same parameter throughout.

\subsection{Damping and Decoherence Functionals}

\paragraph{Damping function.}
The accumulated damping is
\begin{equation}
    \Gamma_{\mathrm{damp}}(t)
    = \int_0^t ds\,\eta(s)
    \propto t^{3-2d_{\mathcal{U}}},
    \quad d_{\mathcal{U}} \neq 3/2,
    \label{eq:damp_func}
\end{equation}
with logarithmic growth at the marginal case
$d_{\mathcal{U}} = 3/2$.

\paragraph{Decoherence functional.}
Coherence between position eigenstates $|x\rangle$ and
$|x'\rangle$ decays as
$|\rho(x,x';t)| \propto
\exp[-(x-x')^2 \Gamma_{\mathrm{decoh}}(t)/2\hbar^2]$,
where
\begin{equation}
    \Gamma_{\mathrm{decoh}}(t)
    = \int_0^t ds \int_0^s ds'\,\nu(s-s').
    \label{eq:decoh_func_def}
\end{equation}

Substituting the high-temperature noise kernel
$\nu(u) \propto T\,u^{-(2d_{\mathcal{U}}-3)}$ and
performing the double integral:
\begin{equation}
    \Gamma_{\mathrm{decoh}}(t) \propto t^{5-2d_{\mathcal{U}}},
    \quad d_{\mathcal{U}} \neq 2,\,5/2.
    \label{eq:decoh_power}
\end{equation}

Similarly, in the low-temperature regime, where $\nu(u) \propto u^{-(2d_{\mathcal{U}}-2)}$:

\begin{equation}
    \Gamma_{\mathrm{decoh}}(t) \propto t^{4-2d_{\mathcal{U}}},
    \quad d_{\mathcal{U}} \neq 3/2,\,2.
    \label{eq:decoh_power_lowT}
\end{equation}

Alternatively, via the spectral representation:
\begin{equation}
    \Gamma_{\mathrm{decoh}}(t)
    = \int_0^\infty d\omega\,\frac{J(\omega)}{\omega^2}\,
    \coth\!\left(\tfrac{\beta\omega}{2}\right)
    (1-\cos\omega t),
    \label{eq:decoh_spectral}
\end{equation}
which is UV-finite for $J(\omega) \propto \omega^s$ with
$s < 1$ and provides the correct regularization for
super-Ohmic cases.

\begin{figure}
    \centering
    \includegraphics[width=0.8\linewidth]{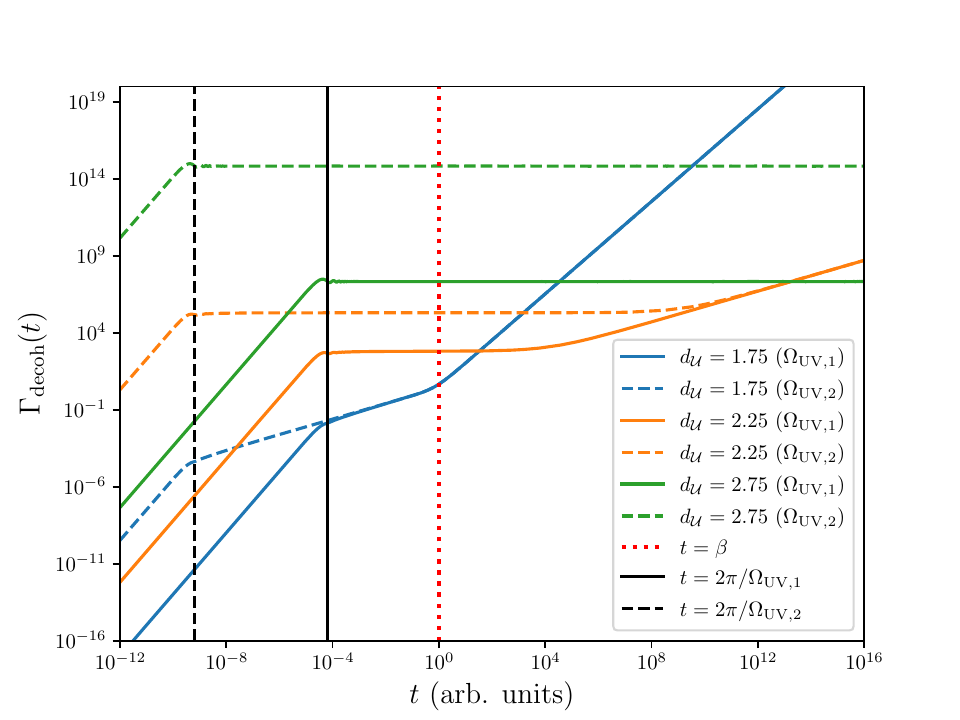}
    \caption{Decoherence functional $\Gamma_{\mathrm{decoh}}(t)$ for the spectral density $J(\omega)=\alpha \omega^s$, $s=2d_\mathcal{U}-3$, $\alpha=1$, with time in units of $\beta$. The plot shows the evolution of the decoherence functional for three different values of $d_\mathcal{U}$ in the sub-Ohmic (blue), super-Ohmic (orange) and ultra super-Ohmic (green) regimes (See Table~\ref{tab:phases}), and two different UV cutoffs $\Omega_{\mathrm{UV},1} = 10^5/\beta$ (solid), $\Omega_{\mathrm{UV},2} = 10^{10}/\beta$ (dashed).}
    \label{fig:decoherence_functional_cutoff_comp}
\end{figure}

Using Eq.~\eqref{eq:decoh_spectral} and introducing a UV cutoff $\Omega_{\mathrm{UV}}$, one can compute the decoherence functional numerically. Fig.~\ref{fig:decoherence_functional_cutoff_comp} shows the result for this calculation in different regimes. In the sub-Ohmic case, we see that the cutoff only affects the functional at times $t \lesssim 2\pi/\Omega_\mathrm{UV}$. Above this time, the functional is well described by the unparticle prediction, the power laws with different exponents in the $t \ll \beta$ and $t \gg \beta$ regimes are visible and match the prediction, and the limit $\Omega_\mathrm{UV} \rightarrow \infty$ is finite and converges to this prediction. This is not the case in the super-Ohmic regime, where the cutoff does impact the decoherence functional at times far above the cutoff scale, as the high-frequency modes in the integral in Eq.~\eqref{eq:decoh_spectral} add a time-independent contribution to the functional, which diverges in the $\Omega_\mathrm{UV} \rightarrow \infty$ limit. If $1<s<2$ ($2<d_\mathcal{U}<5/2$), the power law from the low-frequency modes of the integral eventually overtakes this constant contribution at sufficiently long times, recovering the result from the unparticle description. But if $s>2$ ($d_\mathcal{U}>5/2$), the power law contribution has a negative exponent, so it saturates and never overtakes the constant; thus, coherence survives indefinitely in this regime.

\begin{table}[t]
\centering
\caption{%
Complete set of scaling exponents for a system coupled to an
unparticle bath with dimension $d_{\mathcal{U}}$.
Unless otherwise stated, decoherence and noise exponents
refer to the thermal regime ($t \gg \beta$); the
corresponding vacuum-regime ($t \ll \beta$) exponents are
shifted by $-1$ (see Sec.~\ref{sec:validity}).
All quantities are determined by a single parameter
$d_{\mathcal{U}}$; the consistency relations in the final
row are algebraic identities.
}
\label{tab:exponents}
\begin{tabular}{lccl}
\toprule
\textbf{Quantity} & \textbf{Scaling}
    & \textbf{Exponent} & \textbf{Constraint} \\
\midrule
Spectral density & $J(\omega)\propto\omega^s$
    & $s=2d_{\mathcal{U}}-3$ & --- \\
Dissipation kernel & $\eta(t)\propto t^{-\alpha_\eta}$
    & $\alpha_\eta=2d_{\mathcal{U}}-2$ & $d_{\mathcal{U}}>1$ \\
Noise (high-$T$) & $\nu(t)\propto T\,t^{-\alpha_\nu}$
    & $\alpha_\nu=2d_{\mathcal{U}}-3$ & $d_{\mathcal{U}}>3/2$ \\
Damping & $\Gamma_{\mathrm{damp}}\propto t^{\beta_{\mathrm{damp}}}$
    & $\beta_{\mathrm{damp}}=3-2d_{\mathcal{U}}$ & $d_{\mathcal{U}}\neq3/2$ \\
Decoherence & $\Gamma_{\mathrm{decoh}}\propto t^{\gamma_{\mathrm{decoh}}}$
    & $\gamma_{\mathrm{decoh}}=5-2d_{\mathcal{U}}$ & $d_{\mathcal{U}}\neq2,5/2$ \\
Decoh.\ rate & $\dot\Gamma_{\mathrm{decoh}}\propto t^{\delta_{\mathrm{decoh}}}$
    & $\delta_{\mathrm{decoh}}=4-2d_{\mathcal{U}}$ & --- \\
\midrule
\multicolumn{4}{l}{Consistency relations: $s+\gamma_{\mathrm{decoh}}=2$;\;
$\alpha_\eta+\delta_{\mathrm{decoh}}=2$;\; $\alpha_\nu+\beta_{\mathrm{damp}}=0$.}\\
\bottomrule
\end{tabular}
\end{table}

\subsection{Specific Heat Scaling and Thermodynamic Observables}

The internal energy of the bath follows from the spectral
density $J(\omega) \propto \omega^{2d_{\mathcal{U}}-3}$ and
the Bose-Einstein occupation $n(\omega) =
(e^{\beta\omega}-1)^{-1}$:
\begin{equation}
    U \propto \int_0^\infty d\omega\;
    \omega\,J(\omega)\,n(\omega)
    = \int_0^\infty d\omega\;
    \frac{\omega^{2d_{\mathcal{U}}-2}}{e^{\beta\omega}-1}.
\end{equation}
Here $U$ is the internal energy of the bath degrees of
freedom; in a many-body system at a quantum critical point,
this corresponds to the electronic contribution to the
measured specific heat.
Substituting $u = \beta\omega = \omega/T$:
\begin{equation}
    U \propto T^{2d_{\mathcal{U}}-1}
    \int_0^\infty du\;
    \frac{u^{2d_{\mathcal{U}}-2}}{e^u - 1}
    = T^{2d_{\mathcal{U}}-1}\;\Gamma(2d_{\mathcal{U}}-1)\,
    \zeta(2d_{\mathcal{U}}-1),
\end{equation}
where the integral converges for $d_{\mathcal{U}} > 1$.
Differentiating with respect to $T$:
\begin{equation}
    C(T) = \frac{\partial U}{\partial T}
    \propto T^{2d_{\mathcal{U}}-2},
    \label{eq:cv_scaling}
\end{equation}
or equivalently
\begin{equation}
    \frac{C(T)}{T} \propto T^{2d_{\mathcal{U}}-3}.
    \label{eq:cvT_scaling}
\end{equation}
The extraction formula is therefore
\begin{equation}
    d_{\mathcal{U}} = \frac{x_{C/T}+3}{2},
    \label{eq:dU_cv}
\end{equation}
where $x_{C/T}$ is the power-law exponent of $C/T$ versus
$T$, in direct analogy with the resistivity formula
$d_{\mathcal{U}} = (x_\rho + 2)/2$.

\paragraph{Marginal case and the $-\ln T$ anomaly.}
At $d_{\mathcal{U}} = 3/2$, the exponent in
Eq.~\eqref{eq:cvT_scaling} vanishes: $C/T \propto T^0$.
This is a marginal point in precisely the same sense as the
thermalization transition in the phase diagram
(Table~\ref{tab:phases}, derived in the next subsection):
the leading power-law contribution is constant, and the
first non-trivial correction is logarithmic.
Expanding the integrand near $d_{\mathcal{U}} = 3/2$, the
correction to $U$ at order $(d_{\mathcal{U}}-3/2)$
introduces a factor of $\ln T$, yielding
\begin{equation}
    \left.\frac{C(T)}{T}\right|_{d_{\mathcal{U}}=3/2}
    \propto -\ln\frac{T}{T_0},
    \label{eq:log_cv}
\end{equation}
where $T_0$ is a non-universal scale set by the UV cutoff.

\paragraph{Extraction formulas across channels.}
The following summary collects the $d_{\mathcal{U}}$
extraction formula for each independently measurable
observable channel (see also Table~\ref{tab:exponents}):
\begin{equation}
    d_{\mathcal{U}} = \begin{cases}
        \dfrac{x_\rho + 2}{2}
        & \rho(T) \propto T^{x_\rho}, \\[8pt]
        \dfrac{x_\eta + 2}{2}
        & \eta(T) \propto T^{x_\eta}, \\[8pt]
        \dfrac{x_{C/T} + 3}{2}
        & C(T)/T \propto T^{x_{C/T}}, \\[8pt]
        \dfrac{3}{2}
        & C(T)/T \propto -\ln T \quad\text{(marginal).}
    \end{cases}
    \label{eq:extraction_formulas}
\end{equation}
These formulas are used in Sec.~\ref{sec:strange_metals}
to perform a two-channel consistency check in heavy-fermion
compounds.
All four follow from the single spectral density
$J(\omega) \propto \omega^{2d_{\mathcal{U}}-3}$ via the
appropriate Kubo relation; they are consistency checks,
not independent definitions of $d_{\mathcal{U}}$.

\subsection{Phase Structure}

The three critical dimensions marking qualitative transitions
are:

\paragraph{$d_{\mathcal{U}} = 3/2$ (thermalization
transition).}
The damping exponent $\beta_{\mathrm{damp}} = 3-2d_{\mathcal{U}}$ changes
sign: for $d_{\mathcal{U}} < 3/2$, $\Gamma_{\mathrm{damp}}$
grows (efficient thermalization); for $d_{\mathcal{U}} > 3/2$,
it decays (thermalization fails at long times).
At exactly $d_{\mathcal{U}} = 3/2$, thermalization is logarithmic.

\paragraph{$d_{\mathcal{U}} = 2$ (Ohmic boundary).}
$J(\omega) \propto \omega$; the dissipation kernel
$\eta(t) \propto t^{-2}$ marks the boundary between
non-Markovian ($d_{\mathcal{U}} < 2$, slow kernel decay,
strong memory) and quasi-Markovian
($d_{\mathcal{U}} > 2$, fast decay, finite memory time).
The memory time $\tau_{\mathrm{mem}} =
\int_0^\infty dt\,t\,\eta(t)$ diverges for $d_{\mathcal{U}}
\leq 2$ and is finite for $d_{\mathcal{U}} > 2$.

\paragraph{$d_{\mathcal{U}} = 5/2$ (decoherence phase
transition).}
The decoherence exponent $\gamma_{\mathrm{decoh}} = 5-2d_{\mathcal{U}}$
changes sign.
For $d_{\mathcal{U}} \leq 5/2$: $\Gamma_{\mathrm{decoh}}(t \to \infty) \to
\infty$, coherence is irreversibly lost 
(For $d_{\mathcal{U}} = 5/2$, it grows logarithmically).
For $d_{\mathcal{U}} > 5/2$: $\Gamma_{\mathrm{decoh}}(t \to \infty) \to \mbox{constant}$,
coherence is \emph{protected} at long times.

This last transition---coherence protection by a
super-Ohmic bath---is impossible in any Markovian (Lindblad)
description.
Its physical origin is the extremely short correlation time
of highly super-Ohmic baths: high-frequency modes dominate,
but their rapid oscillations average to zero on the system's
timescale, effectively decoupling the bath at late times.

The complete phase diagram is summarized in
Table~\ref{tab:phases}.

\begin{table}[t]
\centering
\caption{Phase diagram of open-system dynamics as a function
of $d_{\mathcal{U}}$.}
\label{tab:phases}
\begin{tabular}{cllll}
\toprule
$d_{\mathcal{U}}$ & Bath type
    & Thermalization & Decoherence & Memory \\
\midrule
$<3/2$ & Deep sub-Ohmic
    & Power-law & Accelerating & Very strong \\
$=3/2$ & Critical sub-Ohmic
    & Logarithmic & $\propto t^2$ & Strong \\
$3/2$--$2$ & Sub-Ohmic
    & Slowing & Growing & Moderate \\
$=2$ & Ohmic
    & Slow & Linear & Marginal \\
$2$--$5/2$ & Super-Ohmic
    & Fails & Decelerating & Weak \\
$=5/2$ & Decoherence critical
    & Fails & Logarithmic & Very weak \\
$>5/2$ & Ultra super-Ohmic
    & Fails & \emph{Saturates} & Negligible \\
\bottomrule
\end{tabular}
\end{table}

\section{Comparison with Existing Frameworks}
\label{sec:comparison}

\subsection{Caldeira-Leggett Model}

The standard Caldeira-Leggett (CL) master equation for a
particle of mass $m$ in the high-temperature limit is
\begin{equation}
    \frac{d\rho_S}{dt} = -\frac{i}{\hbar}[H_S,\rho_S]
    -\frac{i\gamma}{2\hbar}[x,\{p,\rho_S\}]
    - \frac{\gamma m T}{\hbar^2}[x,[x,\rho_S]],
    \label{eq:CL}
\end{equation}
where $\gamma$ is the damping coefficient of the model.
Here we assume an Ohmic spectral density $J(\omega) = \eta_0\omega$,
where $\eta_0$ is the Ohmic coupling constant, with
high-frequency cutoff, weak coupling, and Markovian dynamics.

The unparticle master equation~\eqref{eq:master_eq}
reduces to Eq.~\eqref{eq:CL} in two limits:

\emph{(i) Ohmic case with coarse-graining.}
For $d_{\mathcal{U}} = 2$, the spectral density is Ohmic.
Coarse-graining over times $t \gg \omega_0^{-1}$ allows
approximating the time-nonlocal integral by an effective
local damping coefficient
$\gamma_{\mathrm{eff}} = \int_0^\infty ds\,\eta(s)$.

\emph{(ii) Large scaling dimension.}
As $d_{\mathcal{U}} \to \infty$,
$\eta(t) \propto t^{-(2d_{\mathcal{U}}-2)}$ decays
infinitely rapidly, approaching $\gamma\,\delta(t-s)$,
exactly reproducing the Markovian limit.

For general $d_{\mathcal{U}}$, the unparticle master equation
constitutes a fractional generalization of the CL equation:
\begin{equation}
    \frac{d\rho_S(t)}{dt} = -\frac{i}{\hbar}[H_S,\rho_S(t)]
    - \frac{C_\alpha}{\hbar^2}
    \int_0^t \frac{ds}{(t-s)^{2d_{\mathcal{U}}-2}}
    [x,[x(s),\rho_S(t)]] + \cdots,
    \label{eq:fractional_CL}
\end{equation}
featuring a Riemann-Liouville fractional integral of order
$\alpha_{\mathrm{dissip}} = 2d_{\mathcal{U}}-2$.
The Caldeira-Leggett model is thus a special case of our formalism, not a
competitor.

\subsection{Lindblad Master Equations}

Lindblad master equations treat decoherence rates as free
parameters and predict exponential decay of coherence,
$|\rho_{\alpha\beta}(t)| \propto e^{-\Gamma_0 t}$,
corresponding to a constant decoherence rate.
Within the unparticle framework, the Lindblad form is
recovered only when $\Gamma_{\mathrm{decoh}}(t) \propto t$,
i.e., $d_{\mathcal{U}} = 2$ (Ohmic).

For all other values of $d_{\mathcal{U}}$, the decoherence
functional is a power law $t^{5-2d_{\mathcal{U}}}$, giving
stretched-exponential rather than purely exponential decay.
Most importantly, for $d_{\mathcal{U}} > 5/2$ the
decoherence functional \emph{saturates} at long times---a
regime entirely inaccessible to Lindblad dynamics and
representing a genuine qualitative difference.

\subsection{Non-Markovian Approaches}

The broader non-Markovian literature~\cite{RevModPhys.89.015001, Breuer:2015zlm}
encompasses many techniques including Nakajima-Zwanzig and time-convolutionless
master equations~\cite{Smirne:2010tzk}, hierarchy of equations of motion~\cite{RevModPhys.89.015001}, and
reaction coordinate methods~\cite{Anto-Sztrikacs:2021ghk}.
The unparticle framework is complementary: it does not
provide a numerical solution method but identifies the
universality class.
Systems with power-law spectral densities previously studied as the spin-boson model~\cite{Leggett1987} are equivalent to the unparticle framework with $d_{\mathcal{U}} = (s+3)/2$ in the limit where the UV cutoff can be taken to infinity, or in situations where it is large compared to all relevant
physical scales. 
In practice, the spin-boson model with a finite UV cutoff
is the more general case; the pure unparticle description
corresponds to the special limit of exact scale invariance
over the full frequency range.
The coherence protection for $s > 2$
(i.e., $d_{\mathcal{U}} > 5/2$) found in
Ref.~\cite{PhysRevA.88.042102} is reproduced here as a consequence
of conformal symmetry rather than a model-specific result.

\subsection{SYK Models and Quantum Chaos}
A natural physical realization of the unparticle framework arises in 
Sachdev--Ye--Kitaev (SYK) models~\cite{Sachdev:1992fk,Kitaev:2015aa}. 
At low energies, the $q$-body SYK Hamiltonian develops emergent conformal 
symmetry with fermion scaling dimension $\Delta = 1/q$~\cite{Maldacena:2016hyu}, 
producing a bath spectral density of the power-law form dictated by the present 
framework, with
\begin{equation}
    d_{\mathcal{U}} = \frac{1}{q} + 1,
\end{equation}
in $(0+1)$ dimensions. Simultaneously, the energy-level statistics of the SYK 
Hamiltonian fall in the GUE universality class of random matrix theory, reflecting 
its maximally chaotic nature and saturation of the Maldacena--Shenker--Stanford 
bound~\cite{Maldacena:2015waa}. An unparticle bath realized by an SYK environment 
therefore admits a twofold description: its two-point function, which drives 
decoherence of the open system, is fixed by conformal symmetry and parametrized 
by $d_{\mathcal{U}}$, while its internal spectral statistics are governed by 
random matrix universality. Whether the decoherence phase transition at 
$d_{\mathcal{U}} = 5/2$ predicted by the present framework corresponds to a 
transition in spectral statistics---such as a crossover from chaotic to integrable 
behavior---in systems where $d_{\mathcal{U}}$ can be tuned through this value 
remains a concrete open problem.

\section{Experimental Validation}
\label{sec:validation_exp}

Our previous work~\cite{companion} provided the primary
empirical validation of the framework: multi-channel transport
data from the unitary Fermi gas yield $d_{\mathcal{U}} = 7/4$
consistently from shear viscosity and thermal conductivity,
two genuinely independent Green's functions, and the
engineered spin-boson experiments of Sun~\emph{et al.}
directly test the consistency relation $s + \gamma_{\mathrm{decoh}} = 2$.
Here, we present a second many-body validation at a different
scaling dimension, establishing that $d_{\mathcal{U}}$
has non-trivial resolution across universality classes.

\subsection{Heavy-Fermion Metals at Quantum Criticality}
\label{sec:strange_metals}

Heavy-fermion compounds tuned to a magnetic quantum critical
point provide a second many-body validation of the unparticle
framework, at a scaling dimension distinct from the unitary
Fermi gas ($d_{\mathcal{U}} = 7/4$).
Unlike the three physical realizations of
Secs.~\ref{sec:ising}--\ref{sec:neutrinos}, where
$d_{\mathcal{U}}$ is derived from CFT operator dimensions
from first principles, the result here is obtained
empirically: two independent transport channels are fit
within the same scale-invariant window, and both yield
$d_{\mathcal{U}} = 3/2$.
This value coincides with the analytical prediction for the
$(1+1)$-dimensional Ising model coupled to the energy
operator (Sec.~\ref{sec:ising}), though the microscopic
identification of the relevant CFT for heavy-fermion quantum
criticality lies beyond the scope of the present work.
The critical sub-Ohmic point $d_{\mathcal{U}} = 3/2$ predicts
linear-in-$T$ resistivity $\rho \propto T$ and logarithmic
specific heat $C/T \propto -\ln T$ as the two leading
observables in the scale-invariant window
(Eq.~\eqref{eq:log_cv} and Table~\ref{tab:exponents}).

\begin{figure*}[t]
\centering

\includegraphics[width=0.7\textwidth]{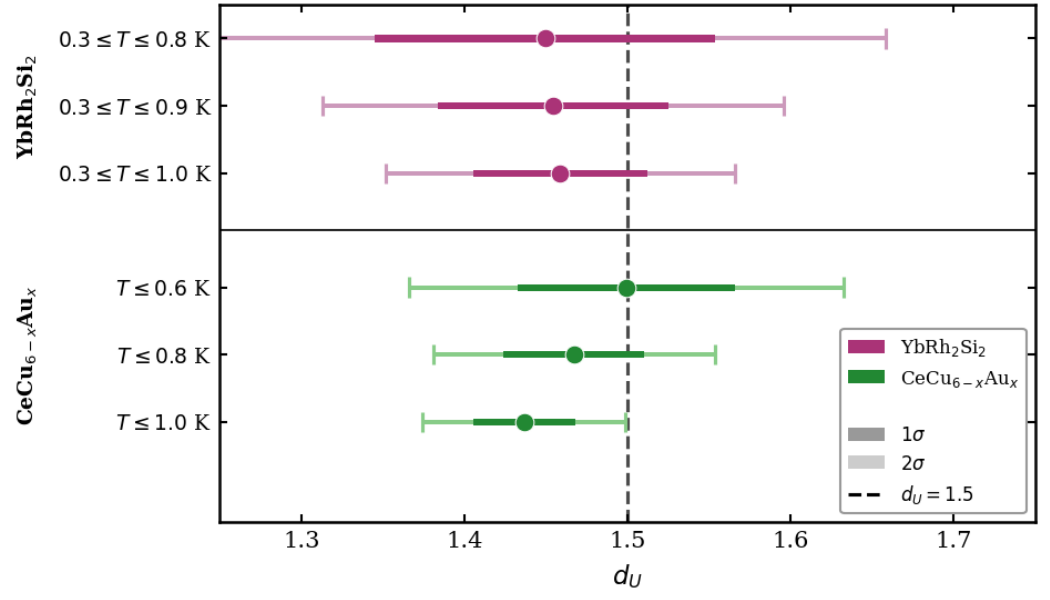}

{\small\linespread{0.8}\selectfont
\textbf{(a)} Extracted $d_{\mathcal U}$ from resistivity fits
$\rho = \rho_0 + AT^x$ with $d_{\mathcal U} = (x+2)/2$,
under progressively lower upper-$T$ cuts.
Both materials converge to $d_{\mathcal U}=3/2$ (dashed line)
at low $T$, where competing effects are suppressed.
Deviations at higher cuts reflect crossover out of the
scale-invariant window (Sec.~\ref{subsec:loopholes}).
Error bars denote $1\sigma$ (inner) and $2\sigma$ (outer).
\par}

\medskip

\includegraphics[width=0.7\textwidth]{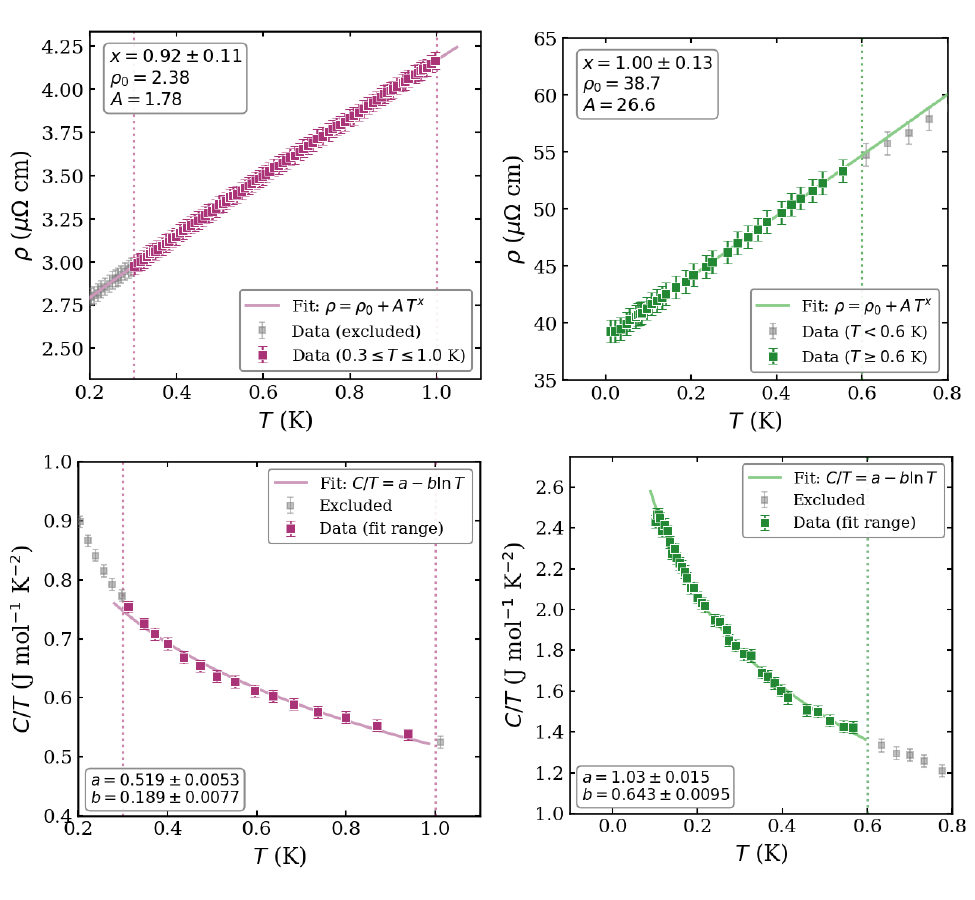}

{\small\linespread{0.8}\selectfont
\textbf{(b)} Two-channel test for
$\mathrm{YbRh_2Si_2}$~\cite{Trovarelli2000} (purple) and
$\mathrm{CeCu_{5.9}Au_{0.1}}$~\cite{Lohneysen1994} (green).
Top: resistivity fits within the scaling window (colored; excluded data grey).
For $\mathrm{YbRh_2Si_2}$, a lower cut $T=0.3\,\mathrm{K}$ removes ordered-phase contamination;
for $\mathrm{CeCu_{5.9}Au_{0.1}}$, tuned to the QCP, only an upper cut
($0.6\,\mathrm{K}$) is applied.
Bottom: $C/T = a - b\ln T$ fits in the same window, consistent with
$d_{\mathcal U}=3/2$.
Agreement between channels constitutes the consistency test.
\par}

\caption{%
\textbf{\small Heavy-fermion validation at $d_{\mathcal U}=3/2$.}
Resistivity and specific heat independently support the same scaling
in two heavy-fermion systems at their quantum critical points,
providing a two-channel check distinct from the unitary Fermi gas
($d_{\mathcal U}=7/4$)~\cite{companion}.
}
\label{fig:strange_metals}
\end{figure*}

\paragraph{Two-channel consistency check.}
We test these predictions against published transport and
thermodynamic data for two materials at their respective
quantum critical points:
$\mathrm{YbRh_2Si_2}$~\cite{Trovarelli2000} and
$\mathrm{CeCu_{5.9}Au_{0.1}}$~\cite{Lohneysen1994}.
For each material we fit the resistivity to
$\rho = \rho_0 + AT^x$ and the specific heat coefficient to
$C/T = a - b\ln T$ within the same scale-invariant
temperature window, extracting $d_{\mathcal{U}} = (x+2)/2$
from the resistivity and verifying that the logarithmic form
holds in the same window.
Results are shown in Fig.~\ref{fig:strange_metals}.

For $\mathrm{YbRh_2Si_2}$, the scale-invariant regime is
$0.3 \leq T \leq 1.0\,\mathrm{K}$; a lower cut is imposed
because this material is close to but not exactly at the
QCP, so ordered-phase behavior appears below
$\sim 0.3\,\mathrm{K}$.
The resistivity fit yields $x = 0.92 \pm 0.11$, giving
$d_{\mathcal{U}} = 1.46 \pm 0.06$, consistent with $3/2$
at the $1\sigma$ level.
The specific heat is logarithmic in the same window
($a = 0.519 \pm 0.005$,
$b = 0.189 \pm 0.008\,\mathrm{J\,mol^{-1}\,K^{-2}}$),
in agreement with the marginal prediction of
Eq.~\eqref{eq:log_cv}.

For $\mathrm{CeCu_{5.9}Au_{0.1}}$, tuned to the QCP at
$T = 0$ with no lower cut required, the scale-invariant
window is $T \leq 0.6\,\mathrm{K}$, above which
Fermi-liquid corrections become visible.
The resistivity yields $x = 1.00 \pm 0.13$, giving
$d_{\mathcal{U}} = 1.50 \pm 0.07$, sitting exactly on the
predicted value.
The specific heat is again logarithmic
($a = 1.03 \pm 0.015$,
$b = 0.643 \pm 0.010\,\mathrm{J\,mol^{-1}\,K^{-2}}$)
in the same window.

In both materials, the resistivity and specific heat channels
are fit independently within the same temperature window and
yield mutually consistent results, constituting a two-channel
test of $d_{\mathcal{U}} = 3/2$ in each system.
The high-$T$ breakdown of the power-law scaling, clearly
visible in Fig.~\ref{fig:strange_metals}, is the expected
crossover out of the scale-invariant window rather than a
failure of the framework: at temperatures where Fermi-liquid
corrections or thermal fluctuations above the QCP regime
become relevant, $J(\omega)$ acquires corrections to the
pure power law, consistent with the loophole analysis of
Sec.~\ref{subsec:loopholes}.

Together with the unitary Fermi gas ($d_{\mathcal{U}} = 7/4$)
from our earlier work, these results demonstrate that
$d_{\mathcal{U}}$ has non-trivial resolution across
universality classes: two distinct values are measured in
three physically unrelated many-body systems, with no shared
fitting parameters.

\section{Quantum Critical Points: The 2D Ising Model}
\label{sec:ising}

\subsection{Critical Theory}

The quantum Ising Hamiltonian
\begin{equation}
    H_{\mathrm{Ising}} = -J\sum_{\langle ij\rangle}
    \sigma_i^x\sigma_j^x - h\sum_i\sigma_i^z
\end{equation}
exhibits a continuous quantum phase transition at $h = h_c$.
At criticality the low-energy effective theory is the
two-dimensional Ising CFT with central charge $c = 1/2$.
This is the minimal unitary CFT and is exactly solvable.
The primary operator spectrum contains three operators:
the identity $\mathds{1}$ ($\Delta = 0$), the spin (order
parameter) $\sigma$ ($\Delta = 1/8$), and the energy density
$\varepsilon$ ($\Delta = 1$).

We consider two spacetime dimensionalities:
the $(1+1)$-dimensional quantum Ising chain
($d_{\mathrm{space}} = 1$) and the $(2+1)$-dimensional
quantum Ising model ($d_{\mathrm{space}} = 2$).
These are \emph{different} conformal field theories.
The $(1+1)$-dimensional chain flows to the 2D Ising CFT,
a rational CFT with central charge $c = 1/2$ and
exactly known operator dimensions~\cite{DiFrancesco1997}:
\begin{equation}
    \Delta_\sigma = \tfrac{1}{8}, \qquad
    \Delta_\varepsilon = 1.
\end{equation}
The $(2+1)$-dimensional model flows instead to the 3D Ising
CFT, a non-rational interacting fixed point with no
closed-form description.
Its scaling dimensions have been determined to high precision
by the conformal bootstrap~\cite{Kos:2016ysd}:
\begin{equation}
    \Delta_\sigma \approx 0.5181, \qquad
    \Delta_\varepsilon \approx 1.4126.
\end{equation}
The two CFTs share the qualitative $\mathbb{Z}_2$-symmetry-breaking
structure of the Ising universality class, but they are
distinct theories with different operator dimensions, which
propagate directly to the unparticle predictions via
Eq.~\eqref{eq:dU_Delta}.
For the $(2+1)$D model coupled to the energy operator,
$\Delta_\varepsilon \approx 1.4126$ and $d = 2$ give
$d_{\mathcal{U}} \approx 1.413$, with spectral exponent
$s \approx -0.17$---close to but distinct from the $1/f$
spectrum, and with decoherence exponent
$\gamma_{\mathrm{decoh}} \approx 2.17$.
\subsection{Setup: Probe Spin}

A probe spin (external qubit or localized impurity) at
position $\mathbf{r}_0$ couples to the critical Ising spins
via
\begin{equation}
    H_{\mathrm{int}} = g\,\sigma_{\mathrm{probe}}^z\,
    \mathcal{O}(\mathbf{r}_0),
\end{equation}
where $\mathcal{O}$ is one of the primary operators of the
Ising CFT.
At criticality, the Ising spins exhibit scale-invariant
fluctuations with no characteristic length or time scale,
forming an unparticle bath for the probe.

\subsection{Spectral Density from CFT Correlators}

The Euclidean two-point function is fixed by conformal symmetry:
\begin{equation}
    \langle \mathcal{O}(\mathbf{x},\tau)\,\mathcal{O}(0,0)
    \rangle_E
    = \frac{C_{\mathcal{O}}}
           {(|\mathbf{x}|^2+\tau^2)^\Delta}.
\end{equation}
After analytic continuation and Fourier transformation at
$\mathbf{k}=0$, the spectral density is (from
Eq.~\eqref{eq:spectral_unique}):
\begin{equation}
    J(\omega) \propto \omega^{2\Delta - d_{\mathrm{space}} - 1},
    \quad \omega > 0.
    \label{eq:ising_spectral}
\end{equation}

\subsection{Results for All Cases}

We apply Eq.~\eqref{eq:ising_spectral} to each operator and
spacetime dimension, then use Eq.~\eqref{eq:dU_Delta} to
extract $d_{\mathcal{U}}$ and Table~\ref{tab:exponents} for
all predictions.
Results are collected in Table~\ref{tab:ising}.

\paragraph{$(1+1)$D, coupling to $\varepsilon$
($\Delta=1$, $d=1$).}
\begin{equation}
    J(\omega) \propto \omega^{2\cdot1-1-1} = \omega^0,
    \quad d_{\mathcal{U}} = \tfrac{3}{2}.
\end{equation}
Flat spectral density; sub-Ohmic.
Predicted decoherence:
$\Gamma_{\mathrm{decoh}}(t) \propto t^2$ (quadratic growth).
This is the marginal thermalization case.

\paragraph{$(2+1)$D, coupling to $\varepsilon$
($\Delta \approx 1.4126$, $d=2$).}

Using the conformal bootstrap value~\cite{Kos:2016ysd}
$\Delta_\varepsilon \approx 1.4126$:
\begin{equation}
    J(\omega) \propto \omega^{2\times1.4126-2-1}
    \approx \omega^{-0.174},
    \quad d_{\mathcal{U}} \approx 1.413.
\end{equation}
The spectral density is close to but distinct from the
$1/f$ spectrum ($s = -1$); the spectral exponent
$s \approx -0.174$ lies between $-1$ and $0$, placing
this case in the deep sub-Ohmic regime but without
the marginal IR behavior of the exact $1/f$ case.
The predicted decoherence exponent is
$\gamma_{\mathrm{decoh}} \approx 5 - 2(1.413) = 2.174$
(slightly super-quadratic growth).
As with the $(1+1)$D spin-operator cases, the mildly
divergent spectral density at $\omega \to 0$ requires
IR regularization at strictly zero temperature; for any
finite $T$ or system size, integrals are finite.

\paragraph{Cases with $d_{\mathcal{U}} < 1$ (spin operator
$\sigma$).}
For coupling to $\sigma$, both $(1+1)$D
($d_{\mathcal{U}} = 5/8$) and $(2+1)$D
($d_{\mathcal{U}} = 1/8$) give $d_{\mathcal{U}} < 1$.
The spectral density diverges as $\omega \to 0$ with exponent
stronger than $\omega^{-1}$, making the Fourier integrals
infrared divergent without a physical cutoff.
This is not a failure of the framework but a physical signal:
the spin operator $\sigma$ is the order-parameter field
at criticality, whose fluctuations are too strongly correlated
to serve as a well-defined dissipative bath at exactly zero
temperature and infinite system size.
In any real system, finite temperature or system size
regularizes the IR, and the unparticle description applies
for $\omega > \omega_{\mathrm{IR}}$.

\begin{table}[t]
\centering
\caption{Unparticle dimensions and scaling exponents for the
quantum Ising model at criticality.
The $(1+1)$D values are exact (2D Ising CFT, $c=1/2$);
the $(2+1)$D values use conformal bootstrap
results~\cite{Kos:2016ysd}.
Cases with $d_{\mathcal{U}} < 1$ require IR regularization.
The $(2+1)$D energy-operator case ($d_{\mathcal{U}} \approx 1.413$, $s \approx -0.17$)
has a mildly divergent spectral density as $\omega \to 0$ and likewise requires
IR regularization at strictly zero temperature and infinite system size;
for any finite $T$ or system size the spectral density is cutoff below
$\omega_{\mathrm{IR}}$ and all integrals are finite.}
\label{tab:ising}
\begin{tabular}{llccccc}
\toprule
\textbf{System} & \textbf{Op.}
    & $\Delta$ & $d_{\mathcal{U}}$
    & $s$ & $\gamma_{\mathrm{decoh}}$ & \textbf{Regime} \\
\midrule
$(1+1)$D & $\sigma$ & $1/8$    & $5/8$  & $-7/4$  & $15/4$
    & IR divergent \\
$(1+1)$D & $\varepsilon$ & $1$ & $3/2$  & $0$     & $2$
    & Sub-Ohmic \\
$(2+1)$D & $\sigma$ & $0.5181$ & $0.518$ & $-1.964$ & $3.964$
    & IR divergent \\
$(2+1)$D & $\varepsilon$ & $1.4126$ & $1.413$ & $-0.174$ & $2.174$
    & Sub-Ohmic \\
\bottomrule
\end{tabular}
\end{table}

\subsection{Experimental Predictions}

\paragraph{Decoherence decay.}
Prepare the probe in $\tfrac{1}{\sqrt{2}}\bigl(\ket{\uparrow} +
\ket{\downarrow}\bigr)$ and measure
$|\rho_{\uparrow\downarrow}(t)|$ via Ramsey interferometry.
Fit to $|\rho_{\uparrow\downarrow}(t)| =
\exp(-C\,t^{\gamma_{\mathrm{decoh}}})$ and extract $\gamma_{\mathrm{decoh}}$.
Predicted values: $\gamma_{\mathrm{decoh}} = 2$ for the $(1+1)$D chain
coupled to $\varepsilon$; $\gamma_{\mathrm{decoh}} \approx 2.174$ for the
$(2+1)$D model (from $\Delta_\varepsilon \approx 1.4126$,
Ref.~\cite{Kos:2016ysd}).

\paragraph{Temperature scaling.}
In the high-$T$ regime, the prefactor $C \propto g^2 T$
(from Eq.~\eqref{eq:nu_highT}).
Varying temperature tests this linear scaling independently
of the exponent measurement.

\paragraph{Consistency check.}
Independently measuring the noise spectrum exponent $s$
and the decoherence exponent $\gamma_{\mathrm{decoh}}$ allows verification
of $s + \gamma_{\mathrm{decoh}} = 2$.
For the $(1+1)$D energy-operator coupling: $s = 0$,
$\gamma_{\mathrm{decoh}} = 2$, so $s + \gamma_{\mathrm{decoh}} = 2$ exactly.

\paragraph{Deviation from criticality.}
Tuning $h$ away from $h_c$ introduces a finite correlation
length $\xi$, setting $\omega_{\mathrm{IR}} \sim v/\xi$.
The power-law decoherence crosses over to exponential decay
at $t \sim \xi/v$.
Observing this crossover and extracting the crossover scale
provides an independent determination of $\xi$.

\paragraph{Experimental platforms.}
The $(1+1)$D chain is accessible in trapped-ion
simulators~\cite{Monroe2021} and Rydberg atom
arrays~\cite{Bernien2017}.
For typical parameters ($v \sim 10^3$ m/s, $g \sim 0.1$,
$T \sim 100\,\mu$K), the decoherence time is of order tens
of milliseconds, well within reach of current experiments.
The $(2+1)$D model is accessible through layered magnetic
materials or nitrogen-vacancy centers near quantum critical
materials.

\section{Inflationary Cosmology}
\label{sec:inflation}
\subsection{Background and Setup}

Inflation provides a natural setting in which to apply the
framework developed above, in a context where scale-invariant
correlators arise from geometry rather than from an interacting
fixed point.
In the de Sitter phase of slow-roll inflation, the scale factor
grows approximately exponentially, $a(t) \simeq e^{Ht}$, and
the de Sitter isometry group $SO(4,1)$ contains dilatations,
enforcing approximate dilation invariance of correlation
functions of light fields in the Bunch-Davies
vacuum~\cite{Bunch:1978yq}.

The uniqueness theorem of Sec.~\ref{sec:uniqueness} is proved
under flat-space conformal invariance and does not strictly apply
in curved spacetime; the de Sitter isometry group $SO(4,1)$
differs from the flat-space conformal group $SO(d+1,1)$, and
the relation~\eqref{eq:dU_Delta} must be rederived in this
geometry.
Nevertheless, the non-Markovian formalism can be applied in
the same spirit, and we show below that doing so reproduces
established results on the quantum-to-classical transition
during inflation~\cite{Kiefer:1998qe, Colas:2024xjy}, identifying
inflation as a particular realization of the framework's phase
structure with $d_{\mathcal{U}} = 2$ defined by matching to
those results.

We model the inflaton $\phi$ (the system) as coupled to an
environment consisting of light scalar fields $\chi$ and/or
gravitons, which acquire scale-invariant correlators from the
de Sitter background.
The interaction takes the form
$\mathcal{L}_{\mathrm{int}} = -(g/2)\phi^2\chi$.

\subsection{De Sitter Correlators and Scaling Dimensions}

For a scalar field of mass $m$ in $(3+1)$-dimensional
de Sitter space, the two-point function in the Bunch-Davies
vacuum is
\begin{equation}
    \langle\chi(x)\chi(0)\rangle_{\mathrm{dS}}
    = \frac{C_\Delta}{(-x^2+i\epsilon)^\Delta},
\end{equation}
with scaling dimension
\begin{equation}
    \Delta = \frac{3}{2} + i\nu,
    \quad
    \nu = \sqrt{\tfrac{9}{4} - \tfrac{m^2}{H^2}}.
    \label{eq:dS_dimension}
\end{equation}
In particular, for a \emph{massless} scalar ($m = 0$):
\begin{equation}
    \Delta_{\mathrm{massless}} = \tfrac{3}{2} + i\tfrac{3}{2}.
\end{equation}
The real part, which governs the spectral density, is
$\mathrm{Re}[\Delta] = 3/2$.
For a massless \emph{graviton}, the transverse-traceless modes
satisfy the same equation of motion as a massless minimally
coupled scalar,\footnote{The transverse-traceless (TT) modes
satisfy $\Box\, h_{ij}^{\mathrm{TT}} = 0$, identical to the
massless scalar equation, giving $m_{\mathrm{eff}}^2 = 0$
and hence $\nu = 3/2$ in Eq.~\eqref{eq:dS_dimension}.}
so $\mathrm{Re}[\Delta_h] = 3/2$ likewise.

\subsection{Spectral Density and Unparticle Dimension}

The spectral density relevant to inflaton decoherence cannot
be read off from the flat-space relation
Eq.~\eqref{eq:dU_Delta}, because the de Sitter isometry group
differs from the flat-space conformal group and the resulting
two-point function of bath operators carries different
geometric weight.
A direct derivation of $J(\omega)$ for the Bunch-Davies bath
of light scalars or gravitons along a co-moving worldline
lies beyond the scope of this article; we therefore
proceed by matching the established literature on
inflationary decoherence~\cite{Kiefer:1998qe, Colas:2024xjy},
which finds linear growth of the decoherence functional in
proper time (or in e-folds $N = Ht$):
\begin{equation}
    \Gamma_{\mathrm{decoh}}(t) \propto H t.
    \label{eq:inflation_linear}
\end{equation}
Comparing to the framework's prediction
$\Gamma_{\mathrm{decoh}} \propto t^{5-2d_{\mathcal{U}}}$
in the thermal regime---appropriate here because the
de Sitter Gibbons-Hawking temperature
$T_{\mathrm{GH}} = H/2\pi$ gives
$\beta_{\mathrm{GH}} \sim H^{-1}$, much shorter than the
inflationary decoherence timescale of $\sim 50$--$60$
e-folds, so the cumulative decoherence is firmly in the
regime $t \gg \beta_{\mathrm{GH}}$---fixes
\begin{equation}
  d_{\mathcal{U}} = 2 \quad \text{(Ohmic).}
    \label{eq:dU_inflation}
\end{equation}
We thus identify inflation as the Ohmic boundary
($d_{\mathcal{U}} = 2$) of the framework's phase structure,
where the framework's spectral density would take the form
$J(\omega) \propto \omega$ if its assumptions applied
directly.
This identification is empirical within the family of
inflationary decoherence calculations, not a derivation from
the universality theorem; the latter would require extending
the theorem to de Sitter isometries, which we do not pursue
here.

\subsection{Why de Sitter Is Special}

A key distinction from condensed matter applications is that
de Sitter isometries enforce approximate dilation invariance
even in the thermal (Bunch-Davies) state.
The Gibbons-Hawking temperature $T_{\mathrm{GH}} = H/2\pi$
gives $\beta_{\mathrm{GH}} \sim H^{-1}$, identical to the
fundamental dynamical timescale.
Unlike condensed matter systems where thermal effects at
$t \gg \beta$ break scale invariance, the de Sitter-invariant
correlator maintains its power-law form at all timescales by
symmetry, and the two-point function does not develop an
exponential Boltzmann factor.
Crucially, $\beta_{\mathrm{GH}} \sim H^{-1}$ is much shorter
than the inflationary decoherence timescale of
$\sim 50$--$60$ e-folds, placing the cumulative decoherence
firmly in the thermal regime $t \gg \beta_{\mathrm{GH}}$,
which justifies the use of the thermal-regime exponent
$\gamma_{\mathrm{decoh}} = 5 - 2d_{\mathcal{U}}$ in matching to
Eq.~\eqref{eq:inflation_linear}.

\subsection{Predictions}

\paragraph{Linear decoherence growth.}
The matching result $d_{\mathcal{U}} = 2$ reproduces
linear decoherence growth $\Gamma_{\mathrm{decoh}}(t)
\propto Ht$ (Eq.~\eqref{eq:inflation_linear}), in agreement
with the results of Kiefer and
Polarski~\cite{Kiefer:1998qe} on the quantum-to-classical
transition during inflation and with the systematic
open-EFT treatment of Colas~\emph{et al.}~\cite{Colas:2024xjy},
from which the matching is performed.

\paragraph{Dissipation kernel.}
For $d_{\mathcal{U}} = 2$: $\eta(t) \propto t^{-2}$,
decaying slowly enough to produce non-negligible memory
effects over times $t \lesssim H^{-1}$, but fast enough
that $\int_0^\infty dt\,t\,\eta(t)$ converges---consistent
with the marginal Ohmic case.

\paragraph{Decoherence timescale.}
$\Gamma_{\mathrm{decoh}} \sim 1$ when
$t_{\mathrm{decoh}} \sim (g^2 H)^{-1}$,
or $N_{\mathrm{decoh}} \sim g^{-2}$ e-folds.
For observable CMB modes, which exit the horizon
$\sim 50$--$60$ e-folds before the end of inflation, even
weak couplings $g \sim 0.1$ ensure complete decoherence.

\paragraph{Massive bath and deviations from $d_{\mathcal{U}} = 2$.}
For a massive bath with $m \sim H$, the decoherence
functional is known to grow sub-linearly in
time~\cite{Colas:2024xjy}, implying $\gamma_{\mathrm{decoh}} < 1$ and hence
$d_{\mathcal{U}} > 2$ within the framework's phase structure.
This would delay the quantum-to-classical transition, which could in turn affect the primordial power spectrum, although a detailed characterization of this effect lies beyond the scope of this work.

\begin{table}[t]
\centering
\caption{Mapping between inflationary bath types and
unparticle framework quantities, obtained by matching to
established inflationary decoherence
calculations~\cite{Kiefer:1998qe, Colas:2024xjy}.
The $d_{\mathcal{U}}$ values are not derived from the
flat-space universality theorem but inferred from the
decoherence exponent $\gamma_{\mathrm{decoh}}$ via $d_{\mathcal{U}} =
(5-\gamma_{\mathrm{decoh}})/2$.}
\label{tab:inflation}
\begin{tabular}{lccc}
\toprule
\textbf{Bath} & $d_{\mathcal{U}}$ & $s$ & $\gamma_{\mathrm{decoh}}$ \\
\midrule
Massless scalar ($m=0$) & $2$ & $1$ & $1$ \\
Graviton              & $2$ & $1$ & $1$ \\
Massive scalar ($m \sim H$) & $>2$ & $>1$ & $<1$ \\
\bottomrule
\end{tabular}
\end{table}

\section{High-Energy Astrophysical Neutrinos}
\label{sec:neutrinos}

\subsection{Neutrino Oscillations and Decoherence}

Standard neutrino oscillations are described by the
PMNS mixing matrix.
For two-flavor mixing with angle $\theta$, the vacuum
survival probability is
\begin{equation}
    P(\nu_e\to\nu_e;\,L)
    = 1 - \sin^2(2\theta)\,
    \sin^2\!\left(\frac{\Delta m^2 L}{4E}\right).
\end{equation}
where $\Delta m^2$ is the difference in squared neutrino masses, $E$ is the neutrino energy, and $L$ the distance traveled from production to detection.
Environmental coupling modifies this to
\begin{equation}
    P(\nu_e\to\nu_e;\,L)
    = 1 - \sin^2(2\theta)\,
    \sin^2\!\left(\frac{\Delta m^2 L}{4E}\right)
    e^{-2\Gamma_{\mathrm{decoh}}(L)},
    \label{eq:survival_decoh}
\end{equation}
where $\Gamma_{\mathrm{decoh}}(L)$ is the accumulated
decoherence over baseline $L$.

\subsection{The Unparticle Bath and the Choice of Regime}

We first clarify what bath we have in mind.
The scale-invariant environment relevant to neutrino
propagation is not a Standard Model background.
The CMB ($T \approx 2.725\,\mathrm{K} \approx
0.235\,\mathrm{meV}$~\cite{Fixsen:2009ug}) and the cosmic
neutrino background (predicted $T \approx 1.95\,\mathrm{K}
\approx 0.168\,\mathrm{meV}$~\cite{Kolb:1990vq}) are SM
thermal relics whose interactions with high-energy neutrinos
are described by ordinary electroweak cross sections, not by
a CFT bath; their effects are independent of the unparticle
analysis.
The bath in our framework is a hypothesized
beyond-Standard-Model scale-invariant sector---a hidden CFT,
an unparticle sector, or a gravitational sector if quantum
gravity contributes scale-invariant fluctuations---to which
propagating neutrinos couple via a higher-dimension operator
\begin{equation}
    \mathcal{L}_{\mathrm{int}} \sim
    \frac{g}{M_*^{n}}\,
    \bar\nu\nu\,\mathcal{O}_{\mathcal{U}},
\end{equation}
where $n$ is fixed by requiring the operator to have
dimension four; the specific value of $n$ depends on the
scaling dimension $d_{\mathcal{U}}$ of $\mathcal{O}_{\mathcal{U}}$
and on the UV completion.
The simplest case $n=1$ (dimension-5 operator) is worked
out explicitly in Sec.~\ref{sec:exp-strategy}.
The effective temperature $T_{\mathcal{U}}$ of this hidden
sector is set by its own cosmological or local history and
is, a priori, independent of any SM background temperature.

\paragraph{The correct regime criterion.}
The vacuum-versus-thermal distinction is governed by the
comparison of the neutrino propagation time $t_{\mathrm{prop}}$
to the thermal correlation time of the bath
$\beta_{\mathcal{U}} = 1/T_{\mathcal{U}}$, as established
in Sec.~\ref{sec:validity} and Table~\ref{tab:validity}.
The relevant question is not whether $E \gg T_{\mathcal{U}}$
but whether $t_{\mathrm{prop}} \gg \beta_{\mathcal{U}}$
or $t_{\mathrm{prop}} \ll \beta_{\mathcal{U}}$.

For IceCube astrophysical neutrinos,
$t_{\mathrm{prop}} \sim 10^{17}\,\mathrm{s}$, corresponding
to a frequency scale $t_{\mathrm{prop}}^{-1} \sim
10^{-32}\,\mathrm{eV}$.
The thermal regime therefore applies whenever
$T_{\mathcal{U}} \gg 10^{-32}\,\mathrm{eV}$---a condition
satisfied by any hidden sector that has ever been in thermal
equilibrium, regardless of how cold it is today.
The vacuum regime, by contrast, requires
$T_{\mathcal{U}} \ll 10^{-32}\,\mathrm{eV}$, which is
an extraordinarily cold hidden sector with no natural
cosmological motivation.
The thermal regime is therefore the generic expectation for
astrophysical neutrinos propagating over cosmological
baselines; the vacuum regime is the special case.

This is the opposite of the naive expectation based on
comparing $E \sim \mathrm{TeV}$ to $T_{\mathcal{U}}$:
the neutrino energy is irrelevant to the regime criterion
because it sets the frequency of the bath modes being
sampled, not the timescale over which the bath correlation
function is integrated.
A high-energy neutrino propagating for $10^{17}$ s
accumulates decoherence from many thermal bath fluctuations
even if each individual interaction samples modes at
$\omega \sim E \gg T_{\mathcal{U}}$; what matters is how
many correlation times $\beta_{\mathcal{U}}$ fit within
$t_{\mathrm{prop}}$.

\paragraph{Thermal regime ($t_{\mathrm{prop}} \gg
\beta_{\mathcal{U}}$).}
This is the generic case for IceCube baselines.
The decoherence exponent is
$\gamma_{\mathrm{decoh}} = 5 - 2d_{\mathcal{U}}$, with
coherence-protection threshold at $d_{\mathcal{U}} = 5/2$.
The decoherence functional acquires a prefactor
$\propto T_{\mathcal{U}}$.
The constraint from $\Delta N_{\mathrm{eff}} \lesssim
0.3$~\cite{Planck:2018vyg, Fields:2019pfx} bounds
$T_{\mathcal{U}}$ from above for a hidden sector that
thermalized before BBN, but imposes no lower bound;
$T_{\mathcal{U}}$ can be arbitrarily small and the thermal
regime still applies provided
$T_{\mathcal{U}} \gg t_{\mathrm{prop}}^{-1}$.

\paragraph{Vacuum regime ($t_{\mathrm{prop}} \ll
\beta_{\mathcal{U}}$).}
This applies only for an extremely cold hidden sector with
$T_{\mathcal{U}} \ll 10^{-32}\,\mathrm{eV}$.
The decoherence exponent shifts to
$\gamma_{\mathrm{decoh}} = 4 - 2d_{\mathcal{U}}$, with
coherence-protection threshold at $d_{\mathcal{U}} = 2$.
There is no $T_{\mathcal{U}}$ prefactor; the decoherence
is driven purely by vacuum fluctuations of the bath.
For point sources at known distances, the vacuum regime
could in principle apply to short-baseline experiments
where $t_{\mathrm{prop}} \sim \beta_{\mathcal{U}}$, but
for cosmological baselines it requires implausibly cold
new physics.

Since $T_{\mathcal{U}}$ is unknown, both regimes are
physically open and we develop predictions for each.
The two regimes predict $\gamma_{\mathrm{decoh}}$ values differing by unity
for the same $d_{\mathcal{U}}$, so a sufficiently precise
measurement of the baseline exponent can identify the
operative regime independently of $d_{\mathcal{U}}$ itself.
The unparticle signal is isolated from SM backgrounds not
by its magnitude but by its distinctive baseline dependence
$\propto L^{\gamma_{\mathrm{decoh}}}$ and energy scaling through $\mathcal{A}(E)$,
neither of which has an SM analog.

\subsection{Coupling and Decoherence Rate}

For a neutrino system coupled to a scale-invariant sector
with dimension $d_{\mathcal{U}}$ via a dephasing operator
$A_\nu = \sigma_z$ in the flavor basis, the master equation
is Eq.~\eqref{eq:master_eq}.
The off-diagonal density matrix element evolves as
\begin{equation}
    \rho_{e\mu}(L)
    = \rho_{e\mu}(0)\,e^{-i\Delta m^2 L/4E}\,
    e^{-\Gamma_{\mathrm{decoh}}(L,E)}.
\end{equation}
As established in Sec.~\ref{sec:neutrinos}, the
thermal regime ($t_{\mathrm{prop}} \gg \beta_{\mathcal{U}}$)
is the generic expectation for cosmological baselines.
We therefore lead with the thermal-regime result and treat
the vacuum regime as the special case.

\paragraph{Thermal regime ($t_{\mathrm{prop}} \gg
\beta_{\mathcal{U}}$).}
The noise kernel is $\nu(t) \propto
T_{\mathcal{U}}\,t^{-(2d_{\mathcal{U}}-3)}$ and the
decoherence functional is
\begin{equation}
    \Gamma_{\mathrm{decoh}}(L,E)
    = \mathcal{B}(E,T_{\mathcal{U}})\,L^{5-2d_{\mathcal{U}}},
    \label{eq:nu_decoh_thermal}
\end{equation}
where the prefactor $\mathcal{B}$ carries an explicit factor
of $T_{\mathcal{U}}$ and has mass dimension
$[\mathcal{B}] = [\mathrm{energy}]^{5-2d_{\mathcal{U}}}$.
The baseline exponent $5-2d_{\mathcal{U}}$ is universal,
independent of the UV completion; the coherence-protection
threshold sits at $d_{\mathcal{U}} = 5/2$.
The corresponding rate per unit baseline length,
\begin{equation}
    \frac{d\Gamma_{\mathrm{decoh}}}{dL}
    = (5-2d_{\mathcal{U}})\,\mathcal{B}(E,T_{\mathcal{U}})\,
    L^{4-2d_{\mathcal{U}}},
\end{equation}
is explicitly $L$-dependent---a non-Markovian feature
inaccessible to a Lindblad treatment.

\paragraph{Vacuum regime ($t_{\mathrm{prop}} \ll
\beta_{\mathcal{U}}$).}
For an extremely cold hidden sector, the noise kernel is
$\nu(t) \propto t^{-(2d_{\mathcal{U}}-2)}$ and
\begin{equation}
    \Gamma_{\mathrm{decoh}}(L,E)
    = \mathcal{A}(E)\,L^{4-2d_{\mathcal{U}}},
    \label{eq:nu_decoh_vacuum}
\end{equation}
with no $T_{\mathcal{U}}$ prefactor.
The baseline exponent shifts by $-1$ relative to the
thermal regime, and the coherence-protection threshold
shifts to $d_{\mathcal{U}} = 2$.

\paragraph{General parametrization of the prefactor.}
In both regimes the prefactor has mass dimension fixed by
requiring $\Gamma_{\mathrm{decoh}}$ to be dimensionless.
We parametrize
\begin{equation}
    \mathcal{A}(E) \sim \frac{g^2}{M_*^m}\,E^n,
    \qquad n - m = 4 - 2d_{\mathcal{U}} ,
    \label{eq:A_parametrize}
\end{equation}
\begin{equation}
    \mathcal{B}(E,T_{\mathcal{U}}) \sim
    \frac{g^2\,T_{\mathcal{U}}}{M_*^m}\,E^{n'},
    \qquad n' - m = 4 - 2d_{\mathcal{U}},
    \label{eq:B_parametrize}
\end{equation}
with $g$ dimensionless and $M_*$ the scale of the
underlying physics.
The dimensional constraints leave one combination of
$(n,m)$ or $(n',m)$ free, to be fixed by the UV completion
or extracted from data alongside $d_{\mathcal{U}}$.

\paragraph{Concrete example: dimension-5 operator.}
For the dimension-5 operator
$\mathcal{L}_{\mathrm{int}} \sim (g/M_*)\,\bar\nu\nu\,
\mathcal{O}_{\mathcal{U}}$,
computing the prefactors from the two-point function gives:
\begin{align}
    \mathcal{A}(E) &\sim \frac{g^2}{M_*^2}\,
    E^{6-2d_{\mathcal{U}}}, \\
    \mathcal{B}(E,T_{\mathcal{U}}) &\sim
    \frac{g^2\,T_{\mathcal{U}}}{M_*^2}\,
    E^{6-2d_{\mathcal{U}}}.
\end{align}
The combined experimental signatures are therefore
\begin{equation}
    \Gamma_{\mathrm{decoh}} \sim
    \begin{cases}
    \dfrac{g^2\,T_{\mathcal{U}}}{M_*^2}\,
    E^{6-2d_{\mathcal{U}}}\,L^{5-2d_{\mathcal{U}}}
    & \text{(thermal)}, \\[10pt]
    \dfrac{g^2}{M_*^2}\,
    E^{6-2d_{\mathcal{U}}}\,L^{4-2d_{\mathcal{U}}}
    & \text{(vacuum)}.
    \end{cases}
    \label{eq:nu_full}
\end{equation}
In the thermal regime, the additional factor of
$T_{\mathcal{U}}/E$ relative to the vacuum case reflects
the enhanced bath occupation at finite temperature.
Both regimes produce independently measurable baseline and
energy exponents, providing two handles on $d_{\mathcal{U}}$
in each case.

\subsection{Comparison with Standard Lindblad Treatment}

The standard phenomenological treatment employs a Lindblad
master equation with constant rate $\Gamma_0$, giving
exponential suppression $e^{-\Gamma_0 L}$.
Table~\ref{tab:lindblad_comparison} summarizes the
qualitative differences.

\begin{table}[t]
\centering
\caption{Comparison of Lindblad and unparticle bath
approaches to neutrino decoherence.
Unparticle exponents are given for the thermal regime
($t_{\mathrm{prop}} \gg \beta_{\mathcal{U}}$), which is
the generic expectation for cosmological baselines;
vacuum-regime values are given in parentheses.}
\label{tab:lindblad_comparison}
\begin{tabular}{lll}
\toprule
\textbf{Property} & \textbf{Lindblad}
    & \textbf{Unparticle bath} \\
\midrule
Coherence decay & $e^{-\Gamma_0 L}$
    & $\exp(-\mathcal{B}\,L^{\gamma_{\mathrm{decoh}}})$ \\
Exponent $\gamma_{\mathrm{decoh}}$ & $1$ (fixed)
    & $5-2d_{\mathcal{U}}$ ($4-2d_{\mathcal{U}}$ in vacuum) \\
Rate & Constant & $\propto L^{\gamma_{\mathrm{decoh}}-1}$ \\
Energy scaling & Ad hoc & via $\mathcal{B}(E,T_{\mathcal{U}})$,
    UV-dependent \\
Long-baseline & Always $\to 0$
    & Protected if $d_{\mathcal{U}}>5/2$ ($>2$ in vacuum) \\
Memory & None (Markovian) & Power-law kernel \\
\bottomrule
\end{tabular}
\end{table}

\subsection{Five Dynamical Regimes}

The five regimes as a function of $d_{\mathcal{U}}$ give
qualitatively distinct oscillation patterns in
Eq.~\eqref{eq:survival_decoh}.
We quote the thermal-regime exponent
$\gamma_{\mathrm{decoh}} = 5 - 2d_{\mathcal{U}}$, which is the generic
expectation for cosmological baselines; the vacuum-regime
exponent $\gamma_{\mathrm{decoh}} = 4 - 2d_{\mathcal{U}}$ is obtained by
shifting all thresholds down by $1/2$ in $d_{\mathcal{U}}$,
as noted in parentheses.

\paragraph{$d_{\mathcal{U}} < 3/2$ (deep sub-Ohmic).}
$\gamma_{\mathrm{decoh}} > 2$: decoherence accelerates faster than quadratic,
coherence lost rapidly.

\paragraph{$d_{\mathcal{U}} = 3/2$ (critical sub-Ohmic).}
$\gamma_{\mathrm{decoh}} = 2$: quadratic decoherence growth,
$\Gamma_{\mathrm{decoh}} \propto L^2$.

\paragraph{$3/2 < d_{\mathcal{U}} < 5/2$ (sub- to super-Ohmic).}
$0 < \gamma_{\mathrm{decoh}} < 2$: decoherence grows as a power law with
exponent between zero and two; coherence is lost at long
baselines, more slowly for larger $d_{\mathcal{U}}$.

\paragraph{$d_{\mathcal{U}} = 5/2$ (decoherence critical,
thermal); $d_{\mathcal{U}} = 2$ (decoherence critical,
vacuum).}
$\gamma_{\mathrm{decoh}} = 0$: marginal case; the power-law formula gives
$L^0$ and the actual growth is logarithmic,
$\Gamma \sim \mathcal{B}\,\ln(L/L_0)$.
Coherence survives as a power law:
$|\rho_{e\mu}(L)| \propto L^{-C}$.

\paragraph{$d_{\mathcal{U}} > 5/2$ (super-Ohmic, thermal);
$d_{\mathcal{U}} > 2$ (super-Ohmic, vacuum).}
$\gamma_{\mathrm{decoh}} < 0$: coherence is \emph{protected}
at long distances.
Oscillations survive at arbitrarily large baselines,
a signature impossible in any Lindblad treatment.

\subsection{Experimental Strategy}
\label{sec:exp-strategy}

The primary observable is the joint baseline and energy
dependence of the oscillation suppression.
Using the dimension-5 operator result in the thermal regime
(Eq.~\eqref{eq:nu_full}, upper case) as the default
parametrization, the survival probability takes the form
\begin{equation}
    P(\nu_e\to\nu_e;\,L)
    = 1 - \sin^2(2\theta)\,
    \sin^2\!\left(\frac{\Delta m^2 L}{4E}\right)
    \exp\!\left[
    -C\frac{g^2\,T_{\mathcal{U}}}{M_*^2}\,
    E^{6-2d_{\mathcal{U}}}\,
    L^{5-2d_{\mathcal{U}}}
    \right],
\end{equation}
with $C g^2 T_{\mathcal{U}}/M_*^2$ and $d_{\mathcal{U}}$
as free parameters.
For the vacuum regime (Eq.~\eqref{eq:nu_full}, lower case),
the $E$ exponent remains the same, but the $L$ exponent shifts to
$L^{4-2d_{\mathcal{U}}}$, and there is no $T_{\mathcal{U}}$
prefactor.
In both cases the baseline exponent $\gamma_{\mathrm{decoh}}$ is the primary
observable; the energy exponent and the presence or absence
of a $T_{\mathcal{U}}$ prefactor provide additional handles.

\emph{Test 1: Baseline exponent $\gamma_{\mathrm{decoh}}$.}
Lindblad predicts $\gamma_{\mathrm{decoh}} = 1$.
In the thermal regime, the unparticle framework predicts
$\gamma_{\mathrm{decoh}} = 5 - 2d_{\mathcal{U}}$; measuring $\gamma_{\mathrm{decoh}} \neq 1$
with statistical significance rules out the standard
Markovian description.
The coherence-protection threshold at $d_{\mathcal{U}} = 5/2$
($\gamma_{\mathrm{decoh}} = 0$) is a qualitative target: survival of
oscillations at large baselines with no baseline suppression
would be a striking signature.

\emph{Test 2: Energy dependence.}
At fixed $L$, the suppression scales with energy through
$\mathcal{B}(E, T_{\mathcal{U}})$.
For the dimension-5 operator in the thermal regime this
gives $E^{6-2d_{\mathcal{U}}}$; the energy slope constrains
$d_{\mathcal{U}}$ independently of the baseline measurement.
IceCube's wide energy range (TeV to PeV) provides a lever
arm of three orders of magnitude for this test.
A dedicated analysis is in preparation.

\emph{Test 3: Regime identification.}
The thermal and vacuum regimes predict baseline exponents
$\gamma_{\mathrm{decoh}} = 5-2d_{\mathcal{U}}$ and $4-2d_{\mathcal{U}}$
respectively, differing by unity for the same
$d_{\mathcal{U}}$.
However, they predict the same energy exponent:
$E^{6-2d_{\mathcal{U}}}$.
Jointly fitting both the baseline and energy dependence
therefore determines $d_{\mathcal{U}}$ and identifies the
operative regime simultaneously, without prior knowledge of
$T_{\mathcal{U}}$.
A measurement preferring $\gamma_{\mathrm{decoh}} = 5-2d_{\mathcal{U}}$
over $4-2d_{\mathcal{U}}$ would confirm the thermal regime
and provide an indirect constraint on $T_{\mathcal{U}}$
through the overall amplitude.

\section{Regime of Validity}
\label{sec:validity}

The unparticle framework applies wherever scale invariance
holds.
Here we establish the relevant windows for each physical
system and show that thermal corrections modify amplitudes
but not universal exponents.
A summary of regime assignments for each system is given
in Table~\ref{tab:validity}.

\subsection{The Two Regimes and Their Exponents}

From Eq.~\eqref{eq:nu_exact}, the noise kernel has two
power-law regimes:
\begin{equation}
    \nu(t) \propto \begin{cases}
        t^{-(2d_{\mathcal{U}}-2)}
        & t \ll \beta \quad\text{(vacuum)}, \\
        T\cdot t^{-(2d_{\mathcal{U}}-3)}
        & t \gg \beta \quad\text{(thermal)},
    \end{cases}
\end{equation}
with crossover at $t \sim \beta = 1/T$.
The decoherence exponent accordingly shifts:
$\gamma_{\mathrm{decoh}} = 4-2d_{\mathcal{U}}$ in the vacuum
regime and $5-2d_{\mathcal{U}}$ in the thermal regime.
In both cases $d_{\mathcal{U}}$ is the same parameter;
only the exponent and prefactor change.
Scale invariance is broken in amplitude at the crossover
$t \sim \beta$, but the power-law structure is preserved
on both sides.

\subsection{Condensed Matter}

At $T \sim 100$ K, $\beta \sim 10^{-13}$ s.
Standard decoherence experiments operate at nanosecond
to microsecond timescales, deep in the thermal regime.
The framework applies with the thermal-regime exponent
$\gamma_{\mathrm{decoh}} = 5-2d_{\mathcal{U}}$ and prefactor $\propto T$.

\emph{Crossover test.}
At millikelvin temperatures, observe the transition
from $\gamma_{\mathrm{decoh}} = 4-2d_{\mathcal{U}}$ (vacuum regime,
short times) to $\gamma_{\mathrm{decoh}} = 5-2d_{\mathcal{U}}$
(thermal regime, long times) at $t \sim \beta$
(Sec.~\ref{sec:validity}).
For the $(1+1)$D energy-operator coupling,
$d_{\mathcal{U}} = 3/2$, this predicts a crossover
from $\gamma_{\mathrm{decoh}} = 1$ to $\gamma_{\mathrm{decoh}} = 2$ at $t \sim \beta$.

\subsection{Inflation}

As discussed in Sec.~\ref{sec:inflation}, de Sitter
isometries enforce power-law correlators at all timescales.
The Gibbons-Hawking temperature $T_{\mathrm{GH}} = H/2\pi$
gives $\beta_{\mathrm{GH}} \sim H^{-1}$, identical to the
dynamical timescale.
There is no separation between vacuum and thermal regimes;
instead, de Sitter symmetry guarantees that the correlator
maintains its power-law form throughout.
Inflation is thus the application where the framework's predictions
most cleanly align with established results, although the matching
to those results, rather than a direct derivation from the
universality theorem, is what fixes $d_\mathcal{U} = 2$ in this case.

\subsection{High-Energy Neutrinos}
As established in Sec.~\ref{sec:neutrinos}, the thermal
regime is the generic expectation for IceCube baselines,
since $t_{\mathrm{prop}} \sim 10^{17}\,\mathrm{s} \gg
\beta_{\mathcal{U}}$ for any cosmologically motivated
hidden sector.
The thermal-regime exponent $\gamma_{\mathrm{decoh}} = 5-2d_{\mathcal{U}}$
applies, with the vacuum regime ($\gamma_{\mathrm{decoh}} = 4-2d_{\mathcal{U}}$)
reserved for the special case of an extremely cold hidden
sector with $T_{\mathcal{U}} \ll 10^{-32}\,\mathrm{eV}$.

\begin{table}[t]
\centering
\caption{Regime of validity of the unparticle description
for each physical system.}
\label{tab:validity}
\begin{tabular}{lcccc}
\toprule
\textbf{System} & $T$ & $\beta$
    & $t_{\mathrm{dyn}}$ & \textbf{Regime} \\
\midrule
Cond.\ mat.\ (warm) & $100$ K & $10^{-13}$ s
    & $\gg\beta$ & Thermal \\
Cond.\ mat.\ (cold) & $10$ mK & $10^{-9}$ s
    & $\sim\beta$ & Crossover \\
Inflation & $H/2\pi$ & $H^{-1}$
    & $H^{-1}$ & de Sitter (exact) \\
High-$E$ neutrinos & $T_{\mathcal{U}} \gg t_{\mathrm{prop}}^{-1}$
    & $\beta_{\mathcal{U}}$
    & $10^{17}$ s & Thermal (generic) \\
\bottomrule
\end{tabular}
\end{table}

\section{Experimental Roadmap}
\label{sec:experiments}

\subsection{Trapped-Ion Quantum Simulators}

Trapped-ion systems can simulate the quantum Ising model
with tunable parameters and provide a probe spin weakly
coupled to the critical chain.

\emph{Protocol.}
(1) Tune the transverse field to $h = h_c$ by monitoring
the vanishing of the energy gap.
(2) Prepare the probe in $\tfrac{1}{\sqrt{2}}\bigl(\ket{\uparrow} +
\ket{\downarrow}\bigr)$.
(3) Measure $|\rho_{\uparrow\downarrow}(t)|$ at a series
of times using Ramsey interferometry or spin-echo sequences.
(4) Extract the decoherence exponent $\gamma_{\mathrm{decoh}}$ by fitting
$\ln(-\ln|\rho_{\uparrow\downarrow}|)$ vs $\ln t$.

\emph{Primary prediction.}
$\gamma_{\mathrm{decoh}} = 2$ for $(1+1)$D coupling to $\varepsilon$.

\emph{Consistency test.}
Independently measure the noise spectrum $S(\omega)$
and extract $s$.
Verify $s + \gamma_{\mathrm{decoh}} = 2$.

\emph{Crossover test.}
At millikelvin temperatures, observe the transition
from $\gamma_{\mathrm{decoh}} = 4-2d_{\mathcal{U}}$ (vacuum regime,
short times) to $\gamma_{\mathrm{decoh}} = 5-2d_{\mathcal{U}}$
(thermal regime, long times) at $t \sim \beta$.

\subsection{Neutrino Telescopes}

A dedicated analysis of IceCube data within the unparticle
framework is in preparation and will be described in a future publication.
The primary observables are:

\emph{Baseline and energy dependence.}
Fit the joint baseline and energy dependence of the
oscillation suppression in the parametrization of
Eq.~\eqref{eq:nu_full}.
In the thermal regime (generic for cosmological baselines),
the signature is
$\Gamma_{\mathrm{decoh}} \propto T_{\mathcal{U}}\,
E^{6-2d_{\mathcal{U}}}\,L^{5-2d_{\mathcal{U}}}$;
in the vacuum regime it is
$\Gamma_{\mathrm{decoh}} \propto
E^{6-2d_{\mathcal{U}}}\,L^{4-2d_{\mathcal{U}}}$.
The baseline exponent $\gamma_{\mathrm{decoh}}$ is the primary handle on
$d_{\mathcal{U}}$; the energy exponent and the overall
amplitude carry additional information, including
$T_{\mathcal{U}}$ and $g^2/M_*^2$.

\emph{Regime identification.}
Since the thermal and vacuum regimes predict baseline exponents differing by unity but the same energy exponent, $E^{6-2d_{\mathcal{U}}}$, a joint fit to baseline and energy dependence identifies the operative regime alongside
$d_{\mathcal{U}}$, without prior knowledge of
$T_{\mathcal{U}}$.

\emph{Coherence protection signature.}
For $d_{\mathcal{U}} > 5/2$ (thermal regime) or
$d_{\mathcal{U}} > 2$ (vacuum regime), the oscillation
pattern survives at arbitrarily large baselines.
A non-observation of complete coherence loss at Gpc
baselines would be a qualitative signature of coherence
protection.

Future experiments IceCube-Gen2 and KM3NeT will extend
the accessible energy range and source statistics,
improving sensitivity to $d_{\mathcal{U}}$ by roughly
an order of magnitude.

\subsection{Superconducting Qubits and Engineered Coherence Protection}
\label{sec:qc}

\paragraph{$1/f$ noise in superconducting qubits.}
The dominant decoherence source in superconducting
qubits---transmons, flux qubits, and charge qubits---is
$1/f$ charge and flux noise~\cite{Paladino2014, Krantz2019},
characterized by a noise power spectrum
$S(\omega) \propto 1/\omega$.
In the high-temperature limit, $S(\omega)$ and the bath
spectral density $J(\omega)$ are related by
$S(\omega) \propto T\,J(\omega)/\omega$~\cite{Shnirman_2002},
so $S(\omega) \propto \omega^{-1}$ corresponds to
$J(\omega) \propto \omega^0$, i.e.\ $s = 0$ and
$d_{\mathcal{U}} = 3/2$.
This is precisely the unparticle dimension derived for the
$(1+1)$-dimensional quantum Ising chain coupled to the energy
operator (Sec.~\ref{sec:ising}), providing a field-theoretic
interpretation of $1/f$ noise in superconducting qubits as
the signature of a sub-Ohmic bath at the critical sub-Ohmic
point.

From Table~\ref{tab:exponents}, $d_{\mathcal{U}} = 3/2$
gives a decoherence exponent
\begin{equation}
    \gamma_{\mathrm{decoh}} = 5 - 2d_{\mathcal{U}} = 2,
\end{equation}
predicting quadratic growth
$\Gamma_{\mathrm{decoh}}(t) \propto g^2 T\,t^2$.
This is consistent with existing measurements of qubit
decoherence under $1/f$ noise~\cite{PhysRevLett.97.167001},
and stands in contrast to the linear growth ($\gamma_{\mathrm{decoh}} = 1$)
assumed in standard Lindblad treatments.
The consistency relation $s + \gamma_{\mathrm{decoh}} = 2$
with $s = 0$ fixes $\gamma_{\mathrm{decoh}} = 2$ as a falsifiable prediction,
extractable by fitting the Ramsey coherence envelope
$|\rho_{\uparrow\downarrow}(t)|$ to a stretched exponential
$\exp(-Ct^{\gamma_{\mathrm{decoh}}})$---a minimal modification of the standard
$T_2$ measurement.

\paragraph{Bath-engineered coherence protection.}
The decoherence phase transition at $d_{\mathcal{U}} = 5/2$
(Sec.~\ref{sec:framework}) suggests a qualitative alternative
to active quantum error correction: engineering the bath
spectral density to suppress decoherence passively.
For $d_{\mathcal{U}} > 5/2$, the decoherence functional
acquires a negative exponent and \emph{saturates} at long
times, so that
\begin{equation}
    \lim_{t\to\infty}|\rho_{\uparrow\downarrow}(t)|
    = |\rho_{\uparrow\downarrow}(0)|\,
    e^{-\Gamma_{\mathrm{decoh}}(\infty)},
    \quad
    \Gamma_{\mathrm{decoh}}(\infty) < \infty.
    \label{eq:saturation}
\end{equation}
Coherence is protected by the rapid phase averaging of
high-frequency bath modes, which decouple from the system
at late times---a mechanism absent from any Markovian
description and equivalent to the result of
Leggett~\emph{et al.}~\cite{Leggett1987} in the spin-boson
model, here derived from conformal symmetry.
The tunable parameter is the spectral exponent $s$ of the
electromagnetic or phononic environment seen by the qubit:
engineering $s > 2$ through structured cavities, bandgap
materials, or coupled resonator arrays places the qubit in
the coherence-protected phase.
The saturation~\eqref{eq:saturation} implies that idle
qubit storage becomes asymptotically immune to bath-induced
decoherence, relaxing the requirements on quantum memory
in architectures where idle time between gate operations
is a significant error source.

\section{Summary and Conclusions}
\label{sec:discussion}

We have developed the complete theoretical framework for
open quantum systems coupled to scale-invariant environments,
establishing the following results:

\emph{The uniqueness theorem} (Sec.~\ref{sec:uniqueness})
proves that under locality, Lorentz invariance, unitarity,
and continuous scale invariance, the environment is
necessarily equivalent to an unparticle bath.
Five classes of loophole are identified and treated.
The contrapositive provides a falsifiability criterion:
inconsistent exponents signal a breakdown of scale invariance.

\emph{The complete mathematical framework}
(Sec.~\ref{sec:framework}) provides exact expressions for
all memory kernels including the full vacuum-plus-thermal
decomposition via Matsubara summation, the fractional
Caldeira-Leggett equation valid for arbitrary
$d_{\mathcal{U}}$, and the phase diagram with three
critical dimensions ($3/2$, $2$, $5/2$).

\emph{The Caldeira-Leggett model} (Sec.~\ref{sec:comparison})
is recovered as a special case at $d_{\mathcal{U}} = 2$
in the Markovian limit.
Lindblad master equations are incapable of describing the
decoherence phase transition at $d_{\mathcal{U}} = 5/2$.

\emph{Experimental validation} (Sec.~\ref{sec:validation_exp})
demonstrates that $d_{\mathcal{U}}$ has non-trivial resolution
across universality classes.
Two-channel fits to resistivity and specific heat data for
$\mathrm{YbRh_2Si_2}$~\cite{Trovarelli2000} and
$\mathrm{CeCu_{5.9}Au_{0.1}}$~\cite{Lohneysen1994} at their
respective quantum critical points yield
$d_{\mathcal{U}} = 3/2$ independently from both observables,
consistent with the marginal prediction $C/T \propto -\ln T$
of Eq.~\eqref{eq:log_cv}.
Together with the unitary Fermi gas result
($d_{\mathcal{U}} = 7/4$) from~\cite{companion}, two distinct values of
$d_{\mathcal{U}}$ are now measured across three physically
unrelated many-body systems with no shared fitting parameters.

\emph{Quantum Ising criticality} (Sec.~\ref{sec:ising})
yields $d_{\mathcal{U}} = 3/2$ for the $(1+1)$D chain
coupled to the energy operator, providing a field-theoretic
derivation of $1/f$ noise.
For the $(2+1)$D model, the conformal bootstrap
value~\cite{Kos:2016ysd} $\Delta_\varepsilon \approx 1.4126$
gives $d_{\mathcal{U}} \approx 1.413$ and $\gamma_{\mathrm{decoh}} \approx 2.17$,
close to but distinct from the $1/f$ case.

\emph{Inflationary cosmology} (Sec.~\ref{sec:inflation})
gives $d_{\mathcal{U}} = 2$ for massless scalar and
graviton baths, predicting linear decoherence growth
$\Gamma \propto Ht$ in agreement with the established
inflationary literature.
De Sitter isometries make this the most rigorous
application of the framework.

\emph{High-energy neutrinos} (Sec.~\ref{sec:neutrinos})
generically access the thermal regime of the unparticle bath
owing to their cosmological propagation times, and exhibit
energy- and baseline-dependent decoherence
$\Gamma_{\mathrm{decoh}} \propto \mathcal{B}(E,T_{\mathcal{U}})\,
L^{5-2d_{\mathcal{U}}}$ (thermal regime),
qualitatively distinct from Lindblad predictions.
A dedicated IceCube analysis is in preparation.

\emph{Regime of validity} (Sec.~\ref{sec:validity})
establishes that thermal corrections modify amplitudes
but not universal exponents, and identifies inflation
and high-energy neutrinos as the cleanest applications.

\emph{Experimental roadmap} (Sec.~\ref{sec:experiments}) concretely shows how the framework can be applied in the future in various fields.

The primary experimental handles are the consistency
relations $s + \gamma_{\mathrm{decoh}} = 2$,
$\alpha_\eta + \delta_{\mathrm{decoh}} = 2$, and
$\alpha_\nu + \beta_{\mathrm{damp}} = 0$.
These algebraic identities, not fitting parameters,
provide multi-observable tests of the scale-invariance
hypothesis across all platforms.

\section*{Acknowledgements}
GB acknowledges support from the RCCHU. GB and HS are supported by PID2023-151418NB-I00 funded by MCIU/AEI/10.13039/501100011033/, and by the European ITN project HIDDeN (H2020-MSCA-ITN-2019/860881-HIDDeN).
HS is also supported by the grant FPU23/00257, MCIU. TK is supported by the U.S. Department of Energy, Office of Science, Office of Advanced Scientific Computing Research, Department of Energy Computational Science Graduate Fellowship under Award Number DE-SC0025528. The work of GH was supported by the Neutrino Theory Network Fellowship with contract number 726844, and by the U.S. Department of Energy under award number DE-SC0020262.
CA is supported by the Faculty of Arts and Sciences of Harvard University, the National Science Foundation, the Canadian Institute for Advanced Research, the Research Corporation for Science Advancement, the John Templeton Foundation, and the David \& Lucile Packard Foundation. 

\bibliography{oqs_bib}

\end{document}